\PassOptionsToPackage{prologue,table}{xcolor}
\documentclass[acmtog]{acmart}

\acmSubmissionID{1425}

\gdef\cropInsets{0}

\usepackage{animate}

\usepackage{xargs}
\usepackage[prologue,table]{xcolor}
\usepackage{booktabs} %
\usepackage{tabularx}
\usepackage{caption}
\usepackage{nicematrix}
\usepackage{float}
\usepackage{mathtools}
\usepackage{amsmath,amsfonts}
\usepackage{bm}
\usepackage{microtype}
\usepackage{units}
\usepackage{subcaption}
\usepackage{overpic}
\usepackage{cancel}
\usepackage{wrapfig}
\usepackage{placeins}
\usepackage{empheq}
\usepackage[normalem]{ulem}
\usepackage[export]{adjustbox}
\usepackage[vlined,boxed,ruled]{algorithm2e} %
\usepackage{epigraph}

\SetKw{Or}{or}
\SetKw{And}{and}
\SetStartEndCondition{ }{}{}%
\SetKwProg{Fn}{def}{\string:}{}
\SetKwFunction{Range}{range}%
\SetKw{KwTo}{in}
\SetKwFor{For}{for}{\string:}{}%
\SetKwFor{While}{while}{:}{fintq}%
\SetKwIF{If}{ElseIf}{Else}{if}{:}{elif}{else:}{}%
\SetKwComment{Comment}{\# }{}
\SetKw{Continue}{continue}
\AlgoDontDisplayBlockMarkers
\SetAlgoNoEnd
\SetAlgoNoLine
\DontPrintSemicolon

\SetAlFnt{\small}
\SetAlCapFnt{\small}
\SetAlCapNameFnt{\small}
\SetAlCapHSkip{0pt}
\IncMargin{-\parindent}

\usepackage{xr}
\usepackage{multirow}
\usepackage[capitalize,noabbrev]{cleveref} %
\usepackage{url} %
\usepackage[utf8]{inputenc}

\def\commentType{0}

\ifnum\commentType=0
    \newcommandx{\customComment}[3]{}
    \newcommandx{\customTODO}[3]{}
\fi
\ifnum\commentType=1
    \usepackage[
    type=upperleft,
    unit=in
    ]{fgruler}
    \newcommandx{\customComment}[3]{\textcolor{#2}{\textsl{#1: #3}}}
    \newcommandx{\customTODO}[3]{\textcolor{#2}{\textsl{#1: #3}}}
\fi
\ifnum\commentType=2
    \usepackage[
    type=upperleft,
    unit=in
    ]{fgruler}
    \usepackage{pdfcomment}
    \newcommandx{\customComment}[3]{\pdfcomment[icon=Comment,opacity=0.5,color=#2,author=#1]{#3}}
    \newcommandx{\customTODO}[3]{\pdfcomment[icon=Note,opacity=0.5,color=#2,author=#1]{#3}}
\fi
\ifnum\commentType=3
    \usepackage[
    type=upperleft,
    unit=in
    ]{fgruler}
    \usepackage{pdfcomment} %
    \usepackage{todonotes} %
    \usepackage{silence}
    \newcommandx{\customComment}[3]{\todo[color=#2!40,size=\small]{\textbf{#1:} #3}}
    \newcommandx{\customTODO}[3]{\todo[color=#2!40,size=\small]{\textbf{#1:} #3}}
\fi

\let\originalleft\left %
\let\originalright\right %
\renewcommand{\left}{\mathopen{}\mathclose\bgroup\originalleft} %
\renewcommand{\right}{\aftergroup\egroup\originalright} %

\definecolor{amber}{rgb}{1.0, 0.49, 0.0}
\definecolor{darkgreen}{rgb}{0.0, 0.5, 0.0}
\definecolor{darkblue}{rgb}{0.0, 0.0, 0.5}
\definecolor{darkred}{rgb}{0.5, 0.0, 0.0}
\newcommandx{\All}[1]{\customComment{All}{red}{#1}}
\newcommandx{\Ana}[1]{\customComment{Ana}{amber}{#1}}
\newcommandx{\Justin}[1]{\customComment{Justin}{red}{#1}}
\newcommandx{\justin}[1]{\customComment{Justin}{red}{#1}}
\newcommandx{\vincent}[1]{\customComment{Vincent}{darkblue}{#1}}
\newcommandx{\Oded}[1]{\customComment{Oded}{darkgreen}{#1}}
\newcommandx{\jrk}[1]{\customComment{jrk}{darkgreen}{#1}}

\newcommandx{\Kartik}[1]{\customComment{Oded}{darkred}{#1}}
\newcommandx{\kartik}[1]{\customComment{Oded}{darkred}{#1}}

\newcommandx{\TODO}[1]{\customTODO{TODO}{red}{#1}}
\newcommandx{\todo}[1]{\customTODO{TODO}{red}{#1}}
\newcommandx{\AnaTODO}[1]{\customTODO{Ana TODO}{amber}{#1}}
\newcommandx{\JustinTODO}[1]{\customTODO{Justin TODO}{darkgreen}{#1}}
\newcommandx{\OdedTODO}[1]{\customTODO{Oded TODO}{darkgreen}{#1}}

\newcommand{\REMOVE}[1]{} %

\def\equationautorefname~#1\null{%
  Equation~(#1)\null
}

\newcommand{\vect}[1]{\boldsymbol{#1}}

\newcommand{\tpose}[0]{\top}

\newcommand{\R}[0]{{\mathbb{R}}}

\newcommand{\FS}{\mathbb{F}}
\newcommand{\ES}{\mathbb{E}}
\newcommand{\VS}{\mathbb{V}}

\DeclareMathAlphabet{\mathmybb}{U}{bbold}{m}{n}

\newcommand{\Diff}[1]{\,\mathrm{d}#1}

\newcommand{\Laplacian}[0]{\Delta}

\newcommand{\hodge}{{\star}}

\gdef\useCroppedImages{1}

\usepackage[export]{adjustbox}
\usepackage{calc}
\usepackage{tabularx}
\usepackage{tikz}
\usepackage{xstring}

\newlength{\beautyHeight}
\newlength{\beautyPixWidth}
\newlength{\beautyPixHeight}
\newlength{\insetvsep}
\gdef\useInsetA{0}
\gdef\useInsetB{0}
\gdef\useInsetC{0}

\newcommand{\setInset}[6]{%
    \expandafter\gdef\csname useInset#1\endcsname{1}%
    \expandafter\gdef\csname inset#1Color\endcsname{#2}%
    \expandafter\gdef\csname crop#1X\endcsname{#3}%
    \expandafter\gdef\csname crop#1Y\endcsname{#4}%
    \expandafter\gdef\csname crop#1W\endcsname{#5}%
    \expandafter\gdef\csname crop#1H\endcsname{#6}%
}

\newcommand{\unsetInset}[1]{%
    \expandafter\gdef\csname useInset#1\endcsname{0}%
}

\newcommand{\addBeautyCrop}[8]{%
    \pdfpxdimen=\dimexpr 1 in/72\relax
    \def\beauty{%
        \let\cropR\relax%
        \let\cropB\relax%
        \newlength\cropR%
        \newlength\cropB%
        \setlength\cropR{{#3 px}-{#5 px}-{#7 px}}%
        \setlength\cropB{{#4 px}-{#6 px}-{#8 px}}%
        \sbox0{\includegraphics[width=#2\textwidth,trim={#5px {\cropB} {\cropR} #6px},clip]{#1}}%
        \begin{tikzpicture}
            \node[anchor=north west,inner sep=0] at (0,0) {\usebox0};
            \begin{scope}[x=\wd0/#7, y=\ht0/#8]
            \if\useInsetA1{
                \draw[\insetAColor,very thick] (\cropAX-#5,-\cropAY+#6) rectangle + (\cropAW,-\cropAH);
            }\fi
            \if\useInsetB1{
                \draw[\insetBColor,very thick] (\cropBX-#5,-\cropBY+#6) rectangle + (\cropBW,-\cropBH);
            }\fi
            \if\useInsetC1{
                \draw[\insetCColor,very thick] (\cropCX-#5,-\cropCY+#6) rectangle + (\cropCW,-\cropCH);
            }\fi
            \end{scope}
        \end{tikzpicture}
    }%
    \setlength\beautyHeight{\heightof{\beauty}}%
    \setlength\beautyPixWidth{#3px}%
    \setlength\beautyPixHeight{#4px}%
    \global\beautyHeight=\beautyHeight%
    \global\beautyPixWidth=\beautyPixWidth%
    \global\beautyPixHeight=\beautyPixHeight%
    \begin{adjustbox}{valign=t}
        \beauty{}
    \end{adjustbox}
}

\newcommand{\trimInset}[6]{%
    \let\cropR\relax%
    \let\cropB\relax%
    \newlength\cropR%
    \newlength\cropB%
    \setlength\cropR{{\beautyPixWidth}-{#3 px}-{#5 px}}%
    \setlength\cropB{{\beautyPixHeight}-{#4 px}-{#6 px}}%
    \color{#2}%
    \fbox{\includegraphics[width=1\linewidth,trim={{#3 px} {\cropB} {\cropR} {#4 px}},clip]{#1}}%
}

\newcommand{\addInset}[2]{%
    \color{#2}%
    \fbox{\includegraphics[width=1\linewidth]{#1}}%
}

\newcommand{\auxtimes}{x}
\newcommand{\auxplus}{+}
\newcommand{\auxspace}{ }

\newcommand{\addInsets}[1]{%
    \begin{adjustbox}{valign=t}
        \StrSubstitute{#1}{.}{-}[\baseFileName]
        \begin{adjustbox}{totalheight=1\beautyHeight,tabular={c}}
            \if\useInsetA1%
                \def\cropfile{\baseFileName-\cropAW\auxtimes\cropAH\auxplus\cropAX\auxplus\cropAY-crop}
                \if\cropInsets1
                    \immediate\write18{convert #1 -crop \cropAW\auxtimes\cropAH\auxplus\cropAX\auxplus\cropAY\auxspace -filter point -resize 800\% \cropfile.png}
                \fi
                \if\useCroppedImages1
                    \addInset{\cropfile.png}{\insetAColor}
                \else
                    \trimInset{#1}{\insetAColor}{\cropAX}{\cropAY}{\cropAW}{\cropAH}%
                \fi%
            \fi%
            \if\useInsetB1%
                \if\useInsetA1\\[\insetvsep]\fi%
                \def\cropfile{\baseFileName-\cropBW\auxtimes\cropBH\auxplus\cropBX\auxplus\cropBY-crop}
                \if\cropInsets1
                    \immediate\write18{convert #1 -crop \cropBW\auxtimes\cropBH\auxplus\cropBX\auxplus\cropBY\auxspace -filter point -resize 800\% \cropfile.png}
                \fi
                \if\useCroppedImages1
                    \addInset{\cropfile.png}{\insetBColor}
                \else
                    \trimInset{#1}{\insetBColor}{\cropBX}{\cropBY}{\cropBW}{\cropBH}%
                \fi%
            \fi%
            \if\useInsetC1%
                \if\useInsetB1\\[\insetvsep]\fi%
                \def\cropfile{\baseFileName-\cropCW\auxtimes\cropCH\auxplus\cropCX\auxplus\cropCY-crop}
                \if\cropInsets1
                    \immediate\write18{convert #1 -crop \cropCW\auxtimes\cropCH\auxplus\cropCX\auxplus\cropCY\auxspace -filter point -resize 800\% \cropfile.png}
                \fi
                \if\useCroppedImages1
                    \addInset{\cropfile.png}{\insetCColor}
                \else
                    \trimInset{#1}{\insetCColor}{\cropCX}{\cropCY}{\cropCW}{\cropCH}%
                \fi%
            \fi%
        \end{adjustbox}
    \end{adjustbox}
}

\definecolor{cartoPrismTeal}{rgb}{0.21960784 0.65098039 0.64705882}
\definecolor{cartoPrismOrange}{rgb}{0.88235294 0.48627451 0.01960784}
\definecolor{cartoPrismGreen}{rgb}{0.45098039 0.68627451 0.28235294}
\definecolor{cartoPrismRed}{rgb}{0.8 0.31372549 0.24313725}
\definecolor{cartoPrismPurple}{rgb}{0.58039216 0.20392157 0.43137255}

\definecolor{mathematicaBlue}{rgb}{0.38, 0.51, 0.71}
\definecolor{mathematicaOrange}{rgb}{0.88, 0.61, 0.14}
\definecolor{mathematicaGreen}{rgb}{0.56, 0.69, 0.19}
\definecolor{mathematicaRed}{rgb}{0.92,0.39, 0.21}
\definecolor{mathematicaPurple}{rgb}{0.53, 0.47, 0.7}

\newtheorem*{remark}{Remark}
\usepackage{enumitem,enumerate}
\setlist[itemize]{noitemsep, nolistsep, leftmargin=*}

\usepackage{wrapfig}

\setcopyright{cc}
\setcctype{by-nc-sa}
\acmJournal{TOG}
\acmYear{2025}
\acmVolume{44}
\acmNumber{4}
\acmArticle{}
\acmMonth{8}
\acmPrice{}
\acmDOI{10.1145/3731422}

\begin{document}

\title[Meschers: Geometry Processing of Impossible Objects]{Meschers: Geometry Processing of Impossible Objects}

\author{Ana Dodik}
\email{anadodik@mit.edu}
\orcid{0000-0003-4391-8877}
\affiliation{%
  \institution{MIT CSAIL}
  \country{USA}
}

\author{Isabella Yu}
\orcid{0009-0009-2981-1679}
\affiliation{%
  \institution{MIT CSAIL}
  \country{USA}
}

\author{Kartik Chandra}
\orcid{0000-0002-1835-3707}
\affiliation{%
  \institution{MIT CSAIL}
  \country{USA}
}

\author{Jonathan Ragan-Kelley}
\orcid{0000-0001-6243-9543}
\affiliation{%
  \institution{MIT CSAIL}
  \country{USA}
}

\author{Joshua Tenenbaum}
\orcid{0000-0002-1925-2035}
\affiliation{%
  \institution{MIT}
  \country{USA}
}

\author{Vincent Sitzmann}
\orcid{1234-5678-9012}
\affiliation{%
  \institution{MIT CSAIL}
  \country{USA}
}

\author{Justin Solomon}
\orcid{1234-5678-9012}
\affiliation{%
  \institution{MIT CSAIL}
  \country{USA}
}

\begin{abstract}
  Impossible objects, geometric constructions that humans can perceive but that cannot exist in real life, have been a topic of intrigue in visual arts, perception, and graphics, yet no satisfying computer representation of such objects exists.
  Previous work embeds impossible objects in $3$D, cutting them or twisting/bending them in the depth axis.
  Cutting an impossible object changes its local geometry at the cut, which can hamper downstream graphics applications, such as smoothing, while bending makes it difficult to relight the object.
  Both of these can invalidate geometry operations, such as distance computation.
  As an alternative, we introduce Meschers, meshes capable of representing impossible constructions akin to those found in M.C.\ Escher's woodcuts.
  Our representation has a theoretical foundation in discrete exterior calculus and supports the use-cases above, as we demonstrate in a number of example applications.
  Moreover, because we can do discrete geometry processing on our representation, we can inverse-render impossible objects.
  We also compare our representation to cut and bend representations of impossible objects.
\end{abstract}




\begin{teaserfigure}
  \centering
  \includegraphics[width=\linewidth]{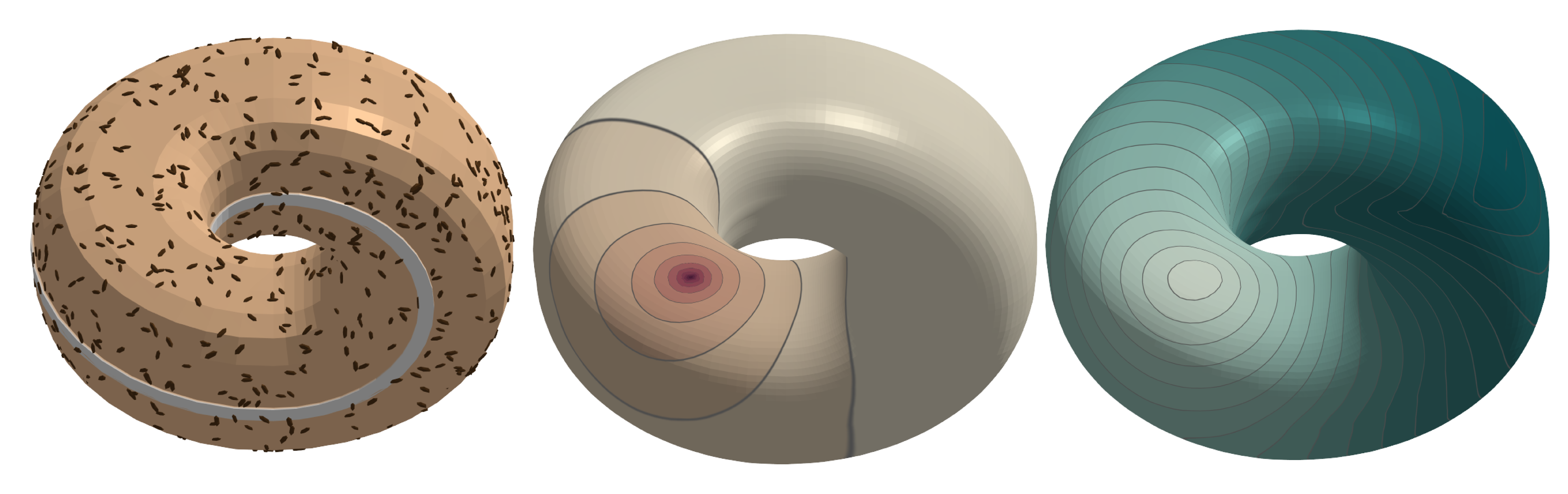}
  \caption{\textsc{The Impossibagel}. The \emph{mescher} is a geometry representation that allows rendering and relighting impossible objects (left), as well as performing intrinsic geometry processing operations like heat diffusion (center) and geodesic distance queries (right).}
  \Description{}\label{fig:teaser}
\end{teaserfigure}  

\maketitle

\section{Introduction}

Recent work in vision science notes that visual art often aligns less with the laws of physics and instead makes sense only in the context of human perception \cite{hertzman2024perspective}.
Extreme examples of this phenomenon include \emph{impossible objects} \citep{penrose1958impossible}, such as those shown in Figure~\ref{fig:teaser}.
Impossible objects often appear in visual art; possibly the most famous examples are the woodcuts of M.C. Escher \shortcite{escher1961graphic}, whereas more recent examples include popular media such as the film \emph{Inception} and videogames \emph{Hocus} and \emph{Monument Valley}.
Computer graphics is the computational language of visual arts, begging the natural question of how to represent constructions like those found in Escher's work computationally.

\begin{figure}[ht]
  \centering
  \includegraphics[width=\linewidth]{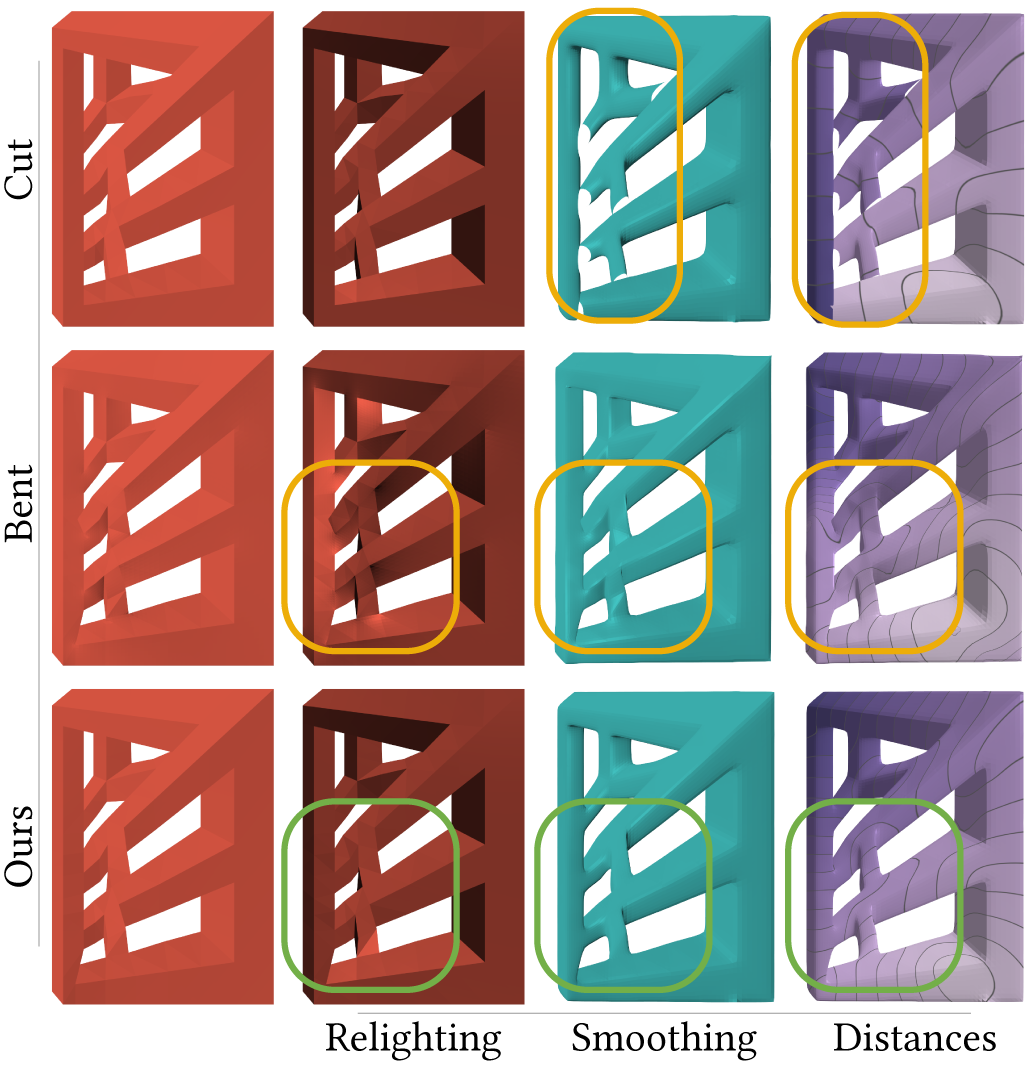}
  \caption{\textsc{Mission: Impossible, Mission: Accomplished.} Here, we compare our method to the two classes of existing methods for rendering impossible geometry. While all three methods support rendering, only our method supports a full set of rendering and geometry processing operations. The \emph{bent} representation leads to artifacts when relighting, due to the perturbed normals. Both representations lead to artifacts when smoothing and when computing geodesic distances.}
  \Description{Comparison to previous methods!}\label{fig:comp}
\end{figure}

At first, it might seem counterintuitive to consider the geometry of impossible constructions, but the impossibility only arises when we attempt to embed impossible objects in $3$D. Focusing
\setlength{\columnsep}{4pt}%
\begin{wrapfigure}{r}{0.3\linewidth}
\begin{center}
\includegraphics[width=\linewidth]{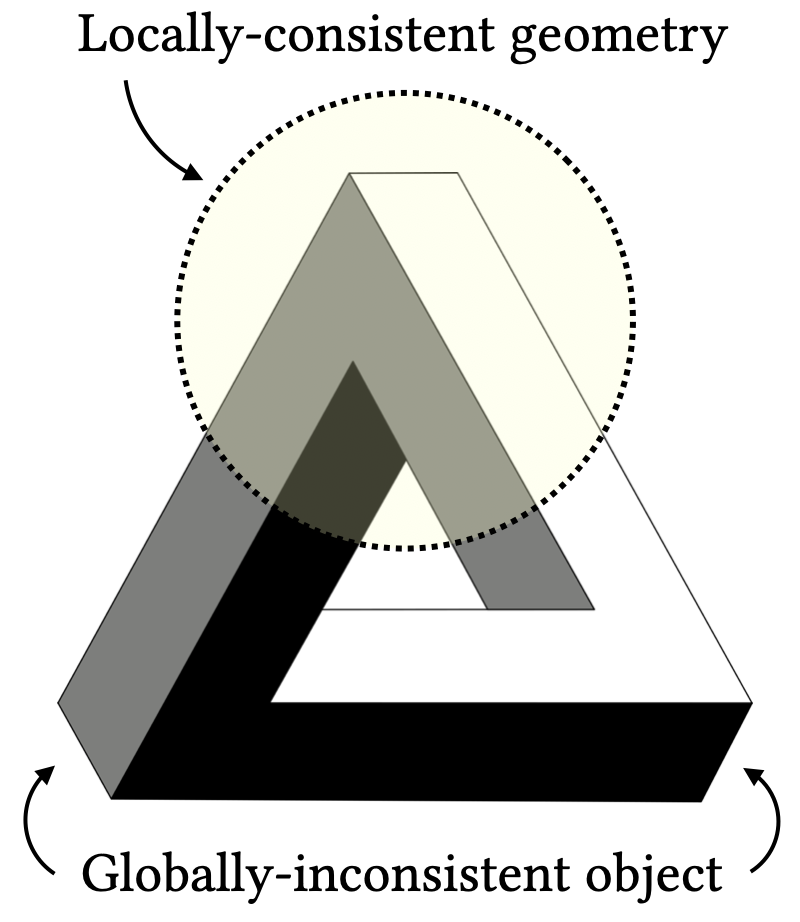}
\end{center}
\end{wrapfigure}
on only one corner of an impossible object, we are able to consistently perceive its \emph{local} geometry.
Its shading offers locally consistent depth and connectivity cues, e.g., a uniform color in a region implies that the region is flat and connected.
It is only when we attempt to assemble---i.e., integrate---local information into a \emph{global} geometry that the impossibility arises~\cite{freud2013b, freud2015, Heinke2021, crane2013deccourse}.
What makes the object impossible is that one cannot assign an absolute depth to every point on its surface without cutting or bending the geometry, which would violate the local depth cues.
In fact, all previous representations of impossible objects resort to either cutting or bending (see Section~\ref{sec:prior-work-rendering}). This often breaks the geometry processing pipeline (Figure~\ref{fig:comp}).

In contrast, we set out to computationally represent impossible objects in a way that still allows us to perform common graphics and geometry processing tasks.
To this end, we introduce the \textbf{mescher}, a perceptually-inspired mesh data structure for Escheresque constructions.
Unlike previous work, our method never integrates the impossible object into $3$D coordinates, side-stepping the need to cut or bend the object (see Section~\ref{sec:prior-work-rendering}).
This means that, unlike in previous work, common graphics operations, such as relighting and intrinsic distance queries, behave consistently with human intuition (Figure~\ref{fig:comp}).

Discrete exterior calculus (DEC) provides a natural language for this task and enables us to construct common discrete differential operators on impossible objects.
In particular, we represent the relative depths of the surface of a mescher using the sum of exact and harmonic primal $1$-forms in DEC.
As a consequence, meschers allow artists to create visuals containing complex impossible objects by relying on graphics tools they are already used to, such as Laplacian smoothing and mesh subdivision.
Our representation also allows us to answer intrinsic shortest path queries, with potential applications in vision science research or in path planning for video games.
Lastly, we demonstrate how the differential operators produced by our representation enable inverse rendering of impossible objects.

In summary, our contributions are:
\begin{itemize}
    \item A geometric representation of impossible objects built on DEC.
    \item The rendering and relighting of impossible objects.
    \item Common modeling operations of smoothing and subdivision.
    \item Answering distance queries on impossible objects.
    \item Inverse rendering of impossible objects.
    \item A thorough qualitative analysis, as well as comparisons with previous methods.
\end{itemize}

\section{Related Work}\label{ss:relatedwork}

For decades, impossible objects have captured the attention of scientists \cite{penrose1958impossible} and artists alike \cite{escher1961graphic, hogarth_satire_1754, reutersvard_triangle_1934}.
Over the years, many attempts were made to mathematically formalize these shapes.
\citet{simon1967information} introduced a model of human perception that involves ``scanning'' a shape to determine whether or not locally plausible-looking regions can be assembled into a globally possible whole.
Similarly, \citet{Gregory1970Eye} hypothesized that humans detect impossibilities through conflicting depth hypotheses, whereas \citet{Draper1978Family} conjectured that the problem involves a violation of the assumption that the object be spatially connected.
In our paper, we phrase all of these observations using the mathematical formalism of \emph{global integrability} (\S\ref{sec:localintegrability}).
We pose our notion of integrability in terms of differential forms, which comprise elements of the de Rham cohomology; note connections between cohomology groups and impossible objects were also explored by \citet{Penrose1993OnTC}.
Lastly, similar to our treatment (\S\ref{sec:localintegrability}, S\ref{ss:order}), \citet{TEROUANNE1984105} attempt to explain perception of impossible objects in terms of a $2$D tesselation of the image and a graph ordering of the image segments.

\subsection{Rendering impossible objects}\label{sec:prior-work-rendering}
Over the past two decades, the computer graphics community has developed many methods for rendering impossible objects. These methods naturally divide into two categories. Methods in the first category (which we call \textbf{``cut representations''}) approach the problem by dissecting an impossible object into a collection of distinct parts, each of which is locally-consistent, but which are assembled into a globally-inconsistent whole \citep{khoh1999animating, savransky1999modeling, owada2008dynafusion, li2024possibleimpossibles, lai20153d}. \citet{taylor2020modeling} extends this approach to compute plausible-looking shadows for impossible objects. The mescher can be seen as the limit of this part-based approach, where each triangle is considered its own ``part.'' This decomposition allows us to perform a variety of powerful intrinsic geometry processing operations on impossible objects.

The second category of methods (which we call \textbf{``bent representations''}) generates view-dependent deformations of a possible object's underlying geometry, such that from a particular non-generic viewpoint the object appears impossible \citep{wu2010modeling, elber2011modeling, sanchez2020make}. These methods differ from ours in that the geometry they \emph{represent} is different from what people \emph{perceive}. For example, while people intuitively see all faces of the Penrose triangle as flat, these methods curve those faces to make them match up correctly in 3D space. Intrinsic geometry processing operations on such representations (e.g.\ computing Gaussian curvature) do not produce intuitive results, as they do on meschers.

In Section~\ref{sec:embedding}, we show how we can recover both of these types of representations from meschers.

\subsection{Modeling and reconstructing impossible objects} How can we produce new impossible objects to render? One approach is to create user interfaces to model impossible objects.
\citet{sugihara1997three} designed a user interface for modeling view-dependent impossible objects, which can then be folded from paper. \citet{owada2008dynafusion}, \citet{inglis2014constructing} and \citet{taylor2020modeling} offer user interfaces for interactively constructing and rendering  complex impossible objects. \citet{li2024possibleimpossibles} procedurally design impossible structures that satisfy simple user-provided geometric constraints. Following this line of work, we allow users to create and edit meschers using operations like subdivision and smoothing.

Another approach to producing impossible objects is to automatically reconstruct their geometry from 2D images. \citet{karpenko2006smoothsketch} attempt to reconstruct a (possible) mesh from a line drawing of an impossible object, but find that the resulting mesh appears ``flattened'' and does not produce the experience of impossibility. As we will show in Section~\ref{ss:inverse}, the mescher enables direct inverse rendering of impossible objects. \citet{weber2023toon3d} reconstruct 3D scenes from geometrically-inconsistent cartoon animations by deforming the input images to maximize consistency. In contrast, the mescher enables recovering inconsistent geometry without warping the input.

\subsection{Perception of impossible objects}

Psychologists and vision scientists have long studied how people perceive impossible objects. The subjective experience of viewing an impossible object is driven by the spatial locality of vision. Because of foveation in the human retina, we perceive the world through small, spatially-localized fixations that we assemble into an abstract global percept \citep{biederman1985human}. Early visual processing extracts local cues like shading and occlusion, which provides the visual system with some information about geometry and depth. This information can be assembled into what \citet{marr2010vision} calls a ``2.5-D sketch'': a precise representation of local surface orientation, combined with a coarse representation of global depth.
Impossible objects surprise us by providing valid local depth cues that fail to integrate into a consistent global depth embedding \citep{freud2013b, freud2015, Heinke2021}. Hence, the mescher represents local geometric properties (e.g.\ orientation, curvature) without committing to global depths.

Importantly, our visual system is robust enough to make partial local sense of such objects \citep{koenderink1998pictorial, linton2023new}---when faced with an impossible object, we only detect the ``impossibility'' when an attempt to integrate multiple fixations leads to an obvious contradiction \citep{schuster1964new}. This feature of human visual perception has important consequences for graphics more broadly \citep{hertzman2024perspective}. For example, classic methods for creating multi-perspective images and panoramas \citep{agrawala2000artistic, agarwala2006photographing, roman2004interactive} rely on people's ability to integrate multiple local perspectives into a coherent global perspective, even when no single global perspective could explain the given image. In this sense, the mescher is a first step towards developing flexible geometric data structures that can represent the wide variety of perspectively-inconsistent scenes that artists wish to depict.

\subsection{Beyond impossible objects} Beyond impossible objects, the graphics community has long developed methods of automatically generating a wide variety of visual illusions \cite{chu2010camouflage, oliva2006hybrid, chi2008self, ma2013change, yu2010embedded, freeman1991motion} and perceptually-ambiguous images \citep{chandra2022designing, burgert2023diffusion, geng2023visual, geng2024factorized, wang2020toward}. These methods are valuable not only for creating interesting and entertaining images, but also for helping understand human visual perception through techniques in visual art \citep{cavanagh2005artist, livingstone2022vision}.

\subsection{Differential mesh representations}

Meschers are closely related to differential representations of meshes \citep{sorkine2006differential}, which have applications in editing and deformation \citep{lipman2004differential, alexa2003differential}. Like meschers, differential representations store a vertex's relationship with its neighbors, rather than its absolute position. A key difference, however, is that meschers only require \emph{local} integrability, not \emph{global}. This is what allows meschers to represent impossible objects.

Our representation is built on machinery of DEC \cite{desbrun2006dec, Hirani2003}, which offers a natural way to talk about differential quantities on meshes; see the survey by \citet{crane2013deccourse} for more information.
Interestingly, this survey in fact alludes to a connection between DEC and impossible objects, without pursuing the topic in more detail \citep[pg 103]{crane2013deccourse}.

DEC and tools from vector field design have been used in other scenarios in need of a local notion of integrability. For example, branched covers of surfaces, which locally integrate to injective parameterizations away from singular points, appear in algorithms for quadrilateral remeshing \cite{kaelberer2007quadcover, bommes2023mixed} and stripe pattern design \cite{Knoppel2015SPS}.

\section{The Mescher}\label{s:mesch}

What makes impossible objects impossible is that they cannot be integrated into $3$D, i.e., we cannot reconstruct \emph{absolute} depth values of the vertices in a way consistent with our depth perception.
Therefore, to model impossible objects in a way that corresponds to perception, we must design a geometry representation that stores topology, screen-space vertex positions, and encodes \emph{relative} depths.

In general, impossible objects assume the existence of an observer coordinate frame and can only be visualized when viewed head-on.
Therefore, we assume an orthographic camera and operate in its coordinate system.
From the perspective of our camera, the $x$ axis is left to right, $y$ down to up, and $z$ backward to forward.

A mescher has the same topology as a typical manifold oriented triangle mesh with (or without) boundary, with sets of triangle faces $\FS$, edges $\ES$, and vertices $\VS$; we assign each edge in $\ES$ and face in $\FS$ an orientation.
To represent a mescher's geometry, we store \emph{per-vertex} screen-space coordinates, $\vect{x}, \vect{y} \in \R^{|\VS|}$.
For depths, we store a value for relative depth \emph{per-edge}, $\vect{\zeta} \in \R^{|\ES|}$.
Intuitively this means that, if an edge $i \in \ES$ connects vertex $p\in\VS$ to vertex $q \in \VS $ then $\zeta_{i}$ represents the \emph{signed} change in depth from $p$ to $q$. The word ``signed'' here signifies that the change from $p$ to $q$ is equal to the negative change from $q$ to $p$.

Looking at this construction from the perspective of discrete exterior calculus (DEC) \cite{Hirani2003, crane2013deccourse, desbrun2006dec}, $\vect{x}$ and $\vect{y}$ are primal $0$-forms (one scalar per vertex), and $\vect{\zeta}$ are primal $1$-forms (one value per edge).
Primal $1$-forms represent the sum total change in a value across an edge; in our case the change in depth.
Leveraging this relationship, we can also build up DEC machinery on a mescher. 
In particular, as the discrete exterior derivatives only depend on the topology, their construction follows in the same way as it does for regular meshes.
This gives us access to discrete exterior derivatives $\Diff_{01} \in \{-1, 0, 1\}^{|\ES|\times|\VS|}$ and $\Diff_{12} \in \{-1, 0, 1\}^{|\FS|\times|\ES|}$ in the primal domain, which map $k$-forms to $(k+1)$-forms.  The transposes of these matrices $\Diff_{12}^\tpose$ and $\Diff_{01}^\tpose$ provide exterior derivatives on the dual mesh.
The remaining operators in the discrete de Rham cohomology, $\hodge_{0}$, $\hodge_{1}$, and $\hodge_{2}$, require some more work to construct from a mescher, as will be discussed in the following section.

\begin{figure}[ht]
  \centering
  \includegraphics[width=\linewidth]{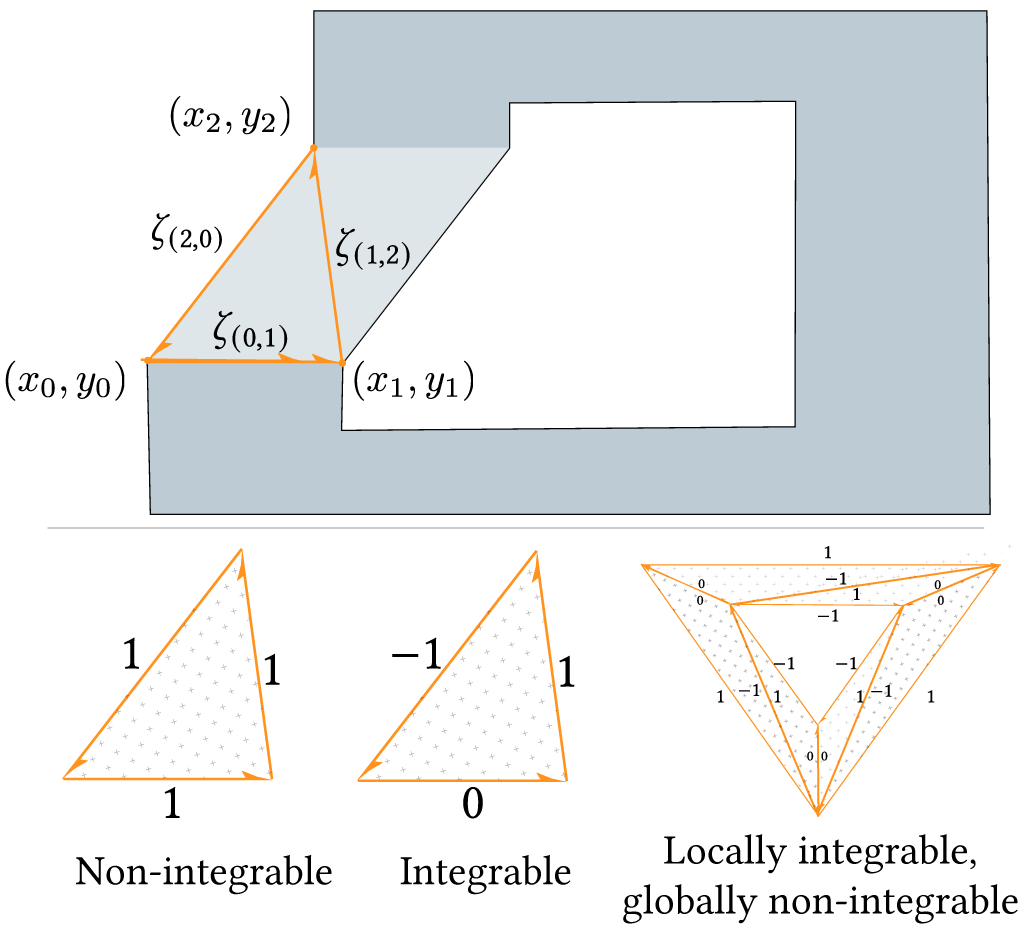}
  \caption{A mescher is defined by a triangle mesh where vertices are stored as a pair of 2D screen-space coordinates $(x, y)$ and edges include a directed relative depth offset $\zeta$. The $\zeta$'s must integrate to zero around individual triangles (orange), but not necessarily around other loops, as shown in the bottom row.}
  \Description{Illustration of representation.}\label{fig:illustration}
\end{figure}

\subsection{Local Integrability and Geometry}\label{sec:localintegrability}

We now articulate the precise space of meschers, which is smaller than the space of all possible discrete $1$-forms.  In particular, consider the example in Figure \ref{fig:illustration}.  Interpreting the $1$-form $\vect{\zeta}$ as a difference in depths per edges of a triangle, the left example is \emph{not integrable}, meaning that the depths do not sum to zero around the cycle of edges in the triangle; the right example is \emph{locally integrable}, meaning they sum to zero.  Local integrability implies that we can locally embed the triangle, i.e., extract a consistent set of depths for the three vertices given the depth of one---a condition that is needed for our mescher to have normal vectors, triangle areas/interior angles, and consequently a cotangent Laplacian operator.  Local integrability does \emph{not} imply global integrability, which is not necessary for a mescher, as shown in the third example; this property distinguishes meschers from globally-embeddable shapes.

In the language of DEC and recalling that $\Diff_{12}$ computes oriented sums of $1$-forms around triangles, the local integrability condition can be expressed as a linear constraint: 
\begin{equation}\label{eq:mescherconstraint}
    \Diff_{12}\vect{\zeta}=\vect{0}.
\end{equation}
This condition is connected to the  discrete Poincar\'e lemma \cite{desbrun2005poincare}.
We will discuss in Section~\ref{ss:obtaining} how we obtain a feasible $\vect{\zeta}$ to initialize our editing system.

Once we have a $\vect{\zeta}$ satisfying the constraint above, we can obtain discrete Hodge star matrices to complete our DEC formulation.  For each triangle on the mesh, we locally integrate $\vect{\zeta}$ to obtain depths for the three vertices (relative to one of the vertices); the screen-space coordinates of the vertices are stored in $\vect{x}$ and $\vect{y}$.  Together, these coordinates allow us to compute triangle areas and half-edge cotangent weights, which we store in diagonal matrices $\star_{0}\in\R^{|\VS|\times|\VS|}$ (one-ring barycentric areas), $\star_{1}\in\R^{|\ES|\times|\ES|}$ (cotangent weights), and $\star_{2}\in\R^{|\FS|\times|\FS|}$ (inverse triangle areas).

With the full set of DEC operators in place, we can use the discrete Hodge decomposition \cite{desbrun2006dec, crane2013deccourse} to characterize the space of mescher $1$-forms.  In particular, any primal $1$-form on a triangle mesh can be written as the sum of a curl-free component, a divergence-free component, and a harmonic component.  The constraint in~\eqref{eq:mescherconstraint} implies that $\vect{\zeta}$ has no divergence-free component, leaving the factorization:
\begin{equation}~\label{eq:hodge}
    \vect{\zeta} = d_{01}\vect{z} + \vect{\omega},
\end{equation}
where $\vect{z}\in\R^{|\VS|}$ and $\vect{\omega}$ is the harmonic component, satisfying $\Diff_{12}\vect{\omega}=\vect{0}$ and $\star_0^{-1}\Diff_{01}^\top\star_1\vect{\omega}=\vect{0}$.

Notice that when there is no harmonic component, i.e.\ when $\vect{\omega}=\vect{0}$, the mescher can be embedded in 3D space, Hence, meschers can represent both possible and impossible objects, and a mescher can be classified as possible or impossible by testing whether $\vect{\omega}=\vect{0}$. %

In contrast to previous work which represents impossible objects through cutting or bending (see Section~\ref{sec:prior-work-rendering}), our representation is compatible with many downstream geometry processing operations (see Section~\ref{s:applications}).
The two discrete differential operators we make use of are the standard 0-form Laplacian, $\Laplacian_{0} \coloneqq \hodge_{0}^{-1} \Diff_{01}^\tpose \hodge_{1} \Diff_{01}$, which use to answer intrinsic distance queries and perform  inverse rendering in Sections~\ref{ss:heat}~and~\ref{ss:inverse}, and the 1-form Hodge Laplacian, $\Laplacian_{1} \coloneqq \hodge_{1}^{-1} \Diff_{12}^\tpose \hodge_{2} \Diff_{12} + \Diff_{01} \hodge_{0}^{-1} \Diff_{01}^\tpose \hodge_{1}$, which can be used for smoothing $1$-forms as in Section~\ref{ss:smooth}.

\begin{remark}[Intrinsic geometry processing]
It is also possible to explain meschers in terms of edge lengths.  The operators above involve only triangle areas and interior angles, all of which are recoverable from the discrete metric.
\end{remark}

\subsection{Finite-element Operators}\label{ss:fem}

While meschers cannot be embedded globally, %
we can construct a \emph{local} coordinate frame for each face.
Differential quantities, such as gradients, only depend on the local tangent spaces, not on the global embedding.
This means that, in addition to DEC, we can also construct standard $1$\textsuperscript{st}-order finite-element operators.
This becomes useful in Section~\ref{ss:heat}, where we need access to per-face gradients of a function.
In our work, we use the gradients, $G_{x}, G_{y}, G_{z} \in \R^{|\FS|\times|\VS|}$, and the diagonal area matrix, $A \in \R^{|\FS|\times |\FS|}$.
As expected, we can verify that we can construct a discrete Laplacian using finite elements, and that $G^\tpose A G = \Diff_{01}^\tpose \star_{1}\Diff_{01}$ and  $A=\star_2^{-1}$.

\begin{figure}[ht]
  \centering
  \includegraphics[width=\linewidth]{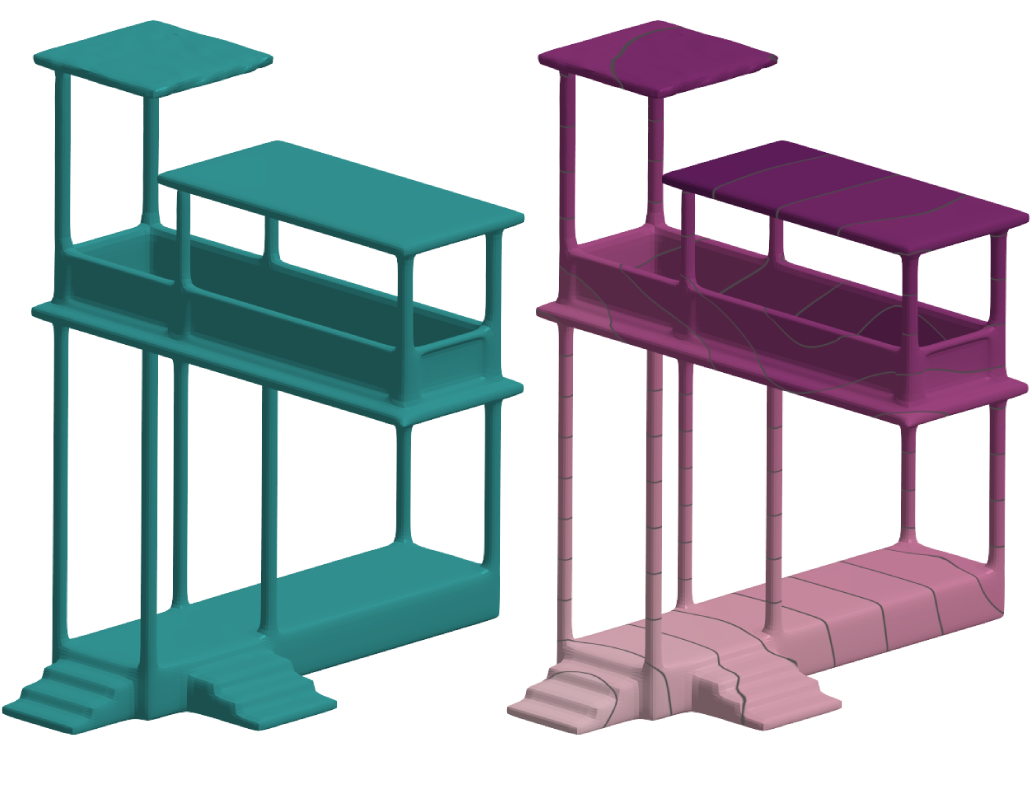}
  \caption{\textsc{Beautiful View}. Using the heat method, we can compute geodesic distances on meschers (Section~\ref{ss:heat}). Here, we show the distance measured from the front steps.}
  \Description{}\label{fig:view}
\end{figure}

\subsection{Obtaining a Mescher}\label{ss:obtaining}

Having explained the inner workings of a mescher, we now briefly discuss how we can obtain one.
Possibly the simplest way is to convert one from a ``cut'' $3$D embedding of the shape, commonly used by previous work (see Section~\ref{sec:prior-work-rendering}).

While impossible surfaces are difficult to reason about, creating a cut representation in $3$D is an artistic practice like any other, with a vibrant community of artists and a plethora of video and textual guides \cite{blenderOutsModelling, sketchfabSpotlightImpossible, elber2011modeling}, books \cite{ernst1986adventures}, as well as online libraries containing hundreds of different impossible objects \cite{impossibleImpossibleWorld, gershonelberEscherReal}.
Other approaches for obtaining impossible objects include direct modeling tools \cite{wu2010modeling}, procedural generation \cite{li2024possibleimpossibles}, or inverse rendering (see \S\ref{ss:inverse}, Figure~\ref{fig:inverse}).

This representation can be modeled using standard software; converting it to a mescher then involves extracting $\vect{\zeta}$ from the $3$D mesh and merging duplicate verticesat the cut as specified by the user.
If the cut $3$D model is imprecise, merging two pairs of vertices that are connected by an edge might lead to a conflict between two different values for $\vect{\zeta}$ along the merged edge.
Simple per-edge merging operations like averaging, however, may yield a $1$-form $\vec{\zeta}$ that does not satisfy Equation~\ref{eq:mescherconstraint}.  This and other similar situations necessitate the need for a means to \emph{project} $\vect{\zeta}$ back onto the feasible set.

Given an unconstrained $\vect{\zeta}^\prime$, we can obtain a $\vect{\zeta}$ consistent with Eq.~\ref{eq:mescherconstraint} by solving the following least-squares problem:
\begin{equation}~\label{eq:meschermin}
\begin{aligned}
    \min_{\vect{\zeta}} \quad & \frac{1}{2} \lVert \vect{\zeta} - \vect{\zeta}^\prime \rVert^2 \\
    \text{s.t.} \quad & \Diff_{12} \vect{\zeta} = \vect{0}.
\end{aligned}
\end{equation}
Introducing a Lagrange multiplier $\vect{\lambda}$, we obtain the optimal $\vect{\zeta}$ by solving a linear system:
\begin{equation}
    \begin{bmatrix}
        I_{|\ES|} & \Diff_{12}^\tpose \\
        \Diff_{12} & 0
    \end{bmatrix}
    \begin{bmatrix}
        \vect{\zeta} \\
        \vect{\lambda}
    \end{bmatrix} =
    \begin{bmatrix}
        \vect{\zeta}^\prime \\
        \vect{0}
    \end{bmatrix}.
\end{equation}
In scenarios where there are multiple conflicting values for each edge's $\zeta$, we minimize its averaged squared difference to both; a straightforward calculation shows that this is identical to solving~\eqref{eq:meschermin} where the corresponding element of $\vect{\zeta}'$ contains the average of the conflicting values.

\begin{remark}
Ideally, in~\eqref{eq:meschermin} the norm $\lVert \vect{\zeta} - \vect{\zeta}^\prime \rVert^2$, would be weighted by $\star_{1}$, i.e., $\lVert \vect{\zeta} - \vect{\zeta}^\prime \rVert^2_\star = (\vect{\zeta} - \vect{\zeta}^\prime)^\tpose \star_{1} (\vect{\zeta} - \vect{\zeta}^\prime)$.
However, there is a ``chicken-and-egg problem:'' In the absence of a consistent $\vect{\zeta}$, we cannot compute a geometric $\star_{1}$.
Therefore, instead of resorting to a complex nonlinear optimization problem, for this projection operator we opt for the simpler solution of using the purely topological $\star_1 \coloneqq \vect{1}$. This is equivalent to performing a discrete Hodge decomposition of $\vect{\zeta}^\prime$ using unit weights for $\star_1$ and setting the divergence-free component to zero.
\end{remark}

Once we have a locally consistent $\vect{\zeta}$, we can render the mescher and compute differential operators, opening doors to various graphics applications (\S\ref{s:applications}).

\subsection{Depth Ordering}\label{ss:order}

\begin{figure}[ht]
  \centering
  \includegraphics[width=\linewidth]{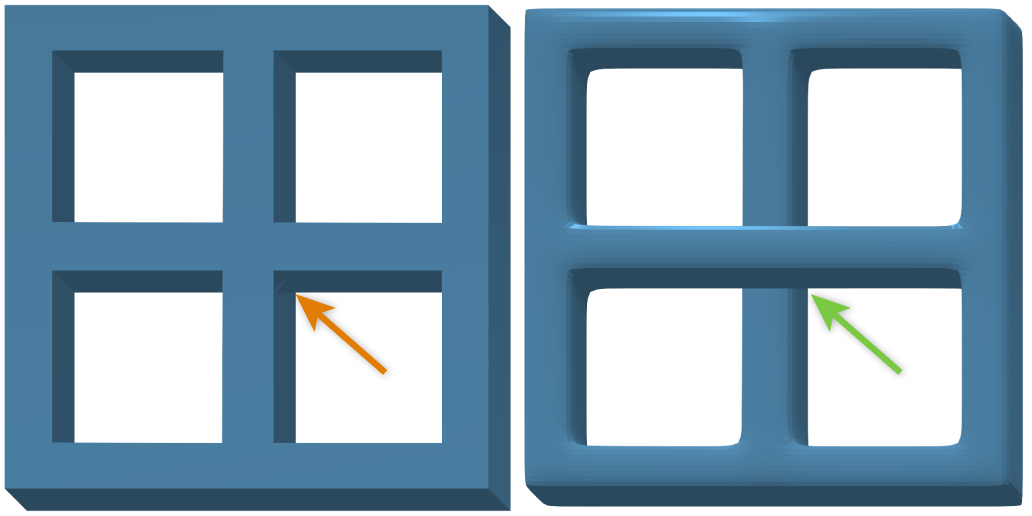}
  \caption{\textsc{Window.} This mescher is globally integrable: na\"ively rendering it makes the bars intersect at the same depth (left). To create an impossibility, an additional form of depth ordering is necessary (Section~\ref{ss:order}). After introducing depth ordering, we can render \textsc{Window} as an impossible object. Here, we further subdivide the mescher and apply our smoothing operator to create specular highlights that emphasize the impossible depth offset.}
  \Description{Window.}\label{fig:window}
\end{figure}

A strong visual cue for depth and connectivity are so-called \emph{T-junctions}.
An example of this is visible in Figure~\ref{fig:window}, where our percept is that the middle part of the horizontal bar is in front of the middle part of the vertical bar, despite the object looking flat otherwise.
So far, we have been dealing with representing the local relative depth ordering of connected geometry, meaning that we require one additional component to our model that would allow us to represent the global depth ordering of disconnected local patches.

We thus represent a \emph{partial} depth ordering to the faces of a mescher  via a directed acyclic graph, where an edge between faces $i$ and $j$ represents that $i$ is perceived to be behind $j$.
This is a separate data structure to the actual mesh, since a graph edge does not imply connectivity or geometry in the same way a mesh edge does.
As the ordering is partial, most faces do not have any depth relationship to eachother, which avoids having to assign an absolute depth order to local patches of an impossible object.

When converting a cut representation into a mescher, we compute intersections between all pairs of triangles and only enforce the ordering for those that do intersect.
To modify the initial ordering, a user can choose two groups of faces and assign one of them to be in front of the other.
We warn the user in case of graph cycles.

It is possible no valid ordering is possible for a given construction.
For example, if the \textsc{Window} mescher in Figure~\ref{fig:window} were represented with few long triangles, a cycle would occur. This problem disappears under refinement, as the ordering is only specified for local patches.
We cannot exclude---but have not observed---pathological examples where refinement might not help.

\section{Operations on Meschers}\label{s:applications}

Having built up the mescher data structure, we can move on to some practical example applications.
We start of with the foundational graphics operations of rendering and lighting (\S\ref{ss:rendering}), and mescher subdivision (\S\ref{ss:subdiv}), before moving on to more complex operations which make use of our differential operators in (\S\ref{ss:heat}-\S\ref{ss:inverse}).

\subsection{Rendering}\label{ss:rendering}

\begin{figure}[ht]
  \centering
  \includegraphics[width=\linewidth]{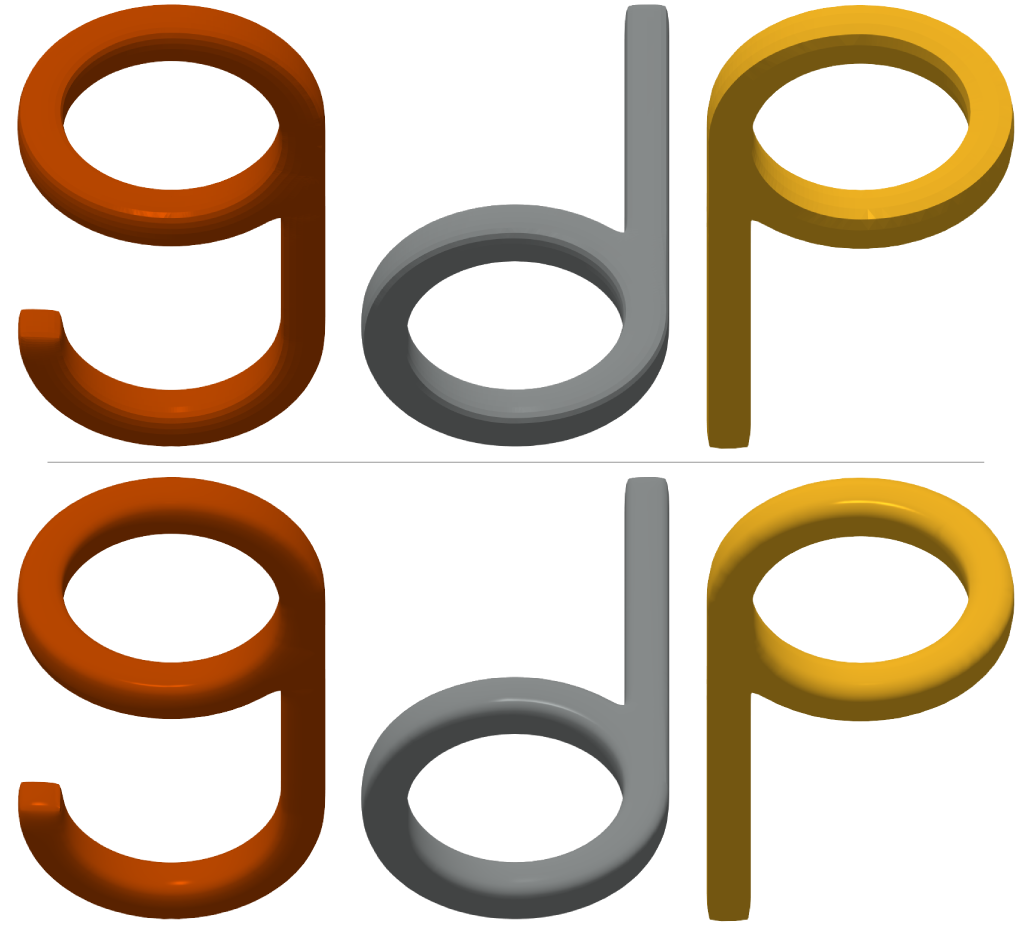}
  \caption{\textsc{Impossible Geometric Data Processing.} Just like in ordinary meshes, meschers can be rendered out with a variety of shading techniques, including both flat (top) and smooth (bottom) shading.}
  \Description{We can render out meschers with smooth shading.}\label{fig:shading}
\end{figure}

To render a mescher, we flatten its triangles by setting their absolute depths to zero and use per-face or per-vertex normals derived from $\vect{\zeta}$ during shading.
Impossible objects are only compatible with directional and environment map illumination.
This is due to the fact that, if the incoming radiance depends only on the direction from which the light is coming, the shading becomes invariant to translations in $\R^3$. More complex emitters such as area lights require knowledge about the absolute $3$D positions of objects and would therefore na\"ively be incompatible with meschers. Meschers are, however, compatible with various local shading models; see Figure~\ref{fig:shading} for a simple example.

For a rendering to respects the depth ordering from Section~\ref{ss:order}, we construct a linear extension from the ordering graph, i.e., a total order that retains the ordering relationships of the partial order.
In practice, this is accomplished by topologically sorting the depth ordering graph.
Once we have a total ordering, we can overlay triangle renders in the order they appear in the topological sort.

\begin{remark}
While our rendering procedure is in essence a normal map, a mescher is not the same as a mesh with a normal map.
One could attempt to represent impossible objects using triangles with arbitrary normals.
However, this would remove the guarantee that neighboring triangles would have the same value of $\zeta$ for their shared edge.
If one were to formulate a least squares problem similar to in Equation~\ref{eq:meschermin} to recover consistent $\zeta$ from arbitrary normals, this would in effect bring us back to the mescher. 
\end{remark}

\begin{figure}[ht]
  \centering
  \includegraphics[width=\linewidth]{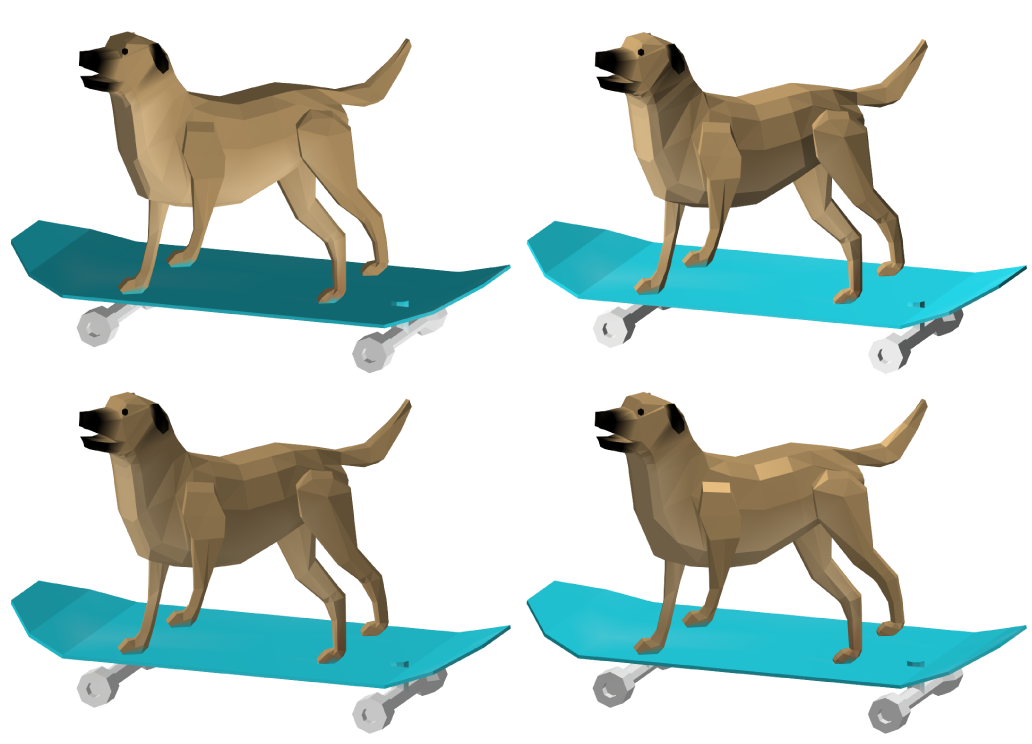}
  \caption{\textsc{Impawssible Dog.} Meschers support relighting. Here, we show the same mescher rendered with four different lighting conditions. This shows that some lighting conditions create a stronger illusory percept than others.}
  \Description{Rendering and relighting.}\label{fig:rendering}
\end{figure}

\subsection{Subdivision}\label{ss:subdiv}

Modeled geometry comes in varying levels of coarseness and further subdivision is often necessary to make them usable with downstream geometry processing tools.
While subdividing $1$-forms on general subdivision surfaces is possible \cite{deGoes2016subdivision}, our reason for subdividing is to make finite elements simulation work and not visual.
Therefore, we opt for the comparably simpler Loop subdivision \cite{Loop1987SmoothSS} and leave other subdivision schemes for future work.

As in Loop subdivision on standard triangle meshes, we begin by introducing a new vertex at the center-point of a triangle's edge, and reconnect the topology to create a total of 4 new triangles.
The $x$ and $y$ positions of the vertices are assigned as expected.
Subdividing the relative depth $\vect{\zeta}$ is possible as we know that the depth increases linearly between two vertices due to our orthographic camera assumption.
Since the 4 new triangles are similar to the original triangle with a scaling factor of one half, a subdivided edge's $\vect{\zeta}$ will equal to the one half of the $\vect{\zeta}$ of the edge to which it is similar.
Subdivision was used to generate virtually all of our results, %
see Figures~\ref{fig:teaser}~\ref{fig:window},~\ref{fig:heart},~\ref{fig:view}, and~\ref{fig:comp}.

\subsection{Heat Diffusion and Geodesics}\label{ss:heat}

A mescher has a discrete metric, enabling intrinsic geometry processing despite not being isometrically embeddable in $\R^3$. A classical intrinsic problem is solving the heat equation on the surface of a shape.
Starting from some initial condition $\vect{u}_0 \in \R^{|\VS|}$, we solve for $\vect{u}$ the linear system given by
\begin{equation}~\label{eq:heat}
    (I + t\Laplacian) \vect{u} = \vect{u}_0,
\end{equation}
where $t$ is the heat diffusion parameter, and $\Laplacian$ the $0$-form Laplacian operator from Section~\ref{sec:localintegrability}.
See Figure~\ref{fig:teaser} for an example of heat diffusion on an impossible object.

Heat diffusion and operations involving the FEM operators in Section~\ref{ss:fem} are sufficient to compute 
for shortest paths on a mescher using the heat method for distance computation \cite{Crane:2017:HMD}.
For a short recap of the method, we first initialize $\vect{u}_0$ to be one at the source vertex and zero elsewhere and compute $\vect{u}$ for a small $t$.
Next, we compute and then normalize the gradient of the diffused function using the FEM gradient operator: $\vect{g} = \frac{G \vect{u}}{\lVert G \vect{u} \rVert_2}$.
To finally obtain the distance $\vect{d}$, we re-integrate the normalized gradient by solving a Poisson equation, $G^TAG \vect{d} = -G^TA \vect{g}$.
We demonstrate intrinsic distance computation on impossible objects in Figures~\ref{fig:teaser},~\ref{fig:view},~\ref{fig:comp}.

\subsection{Smoothing}\label{ss:smooth}

A common task during modeling is to smooth out the rough edges of a modeled mesh.
This can be accomplished via Laplacian smoothing \cite{desbrun2023fairing}.
On embeddable meshes, Laplacian smoothing applies the heat diffusion operator in \eqref{eq:heat} to the $3$D coordinates of the mesh.
In our case, we can apply this strategy to $\vect{x}$ and $\vect{y}$, but it is not obvious how to smooth the $1$-form $\vect{\zeta}$.

\begin{figure}[ht]
  \centering  
  \includegraphics[width=\linewidth]{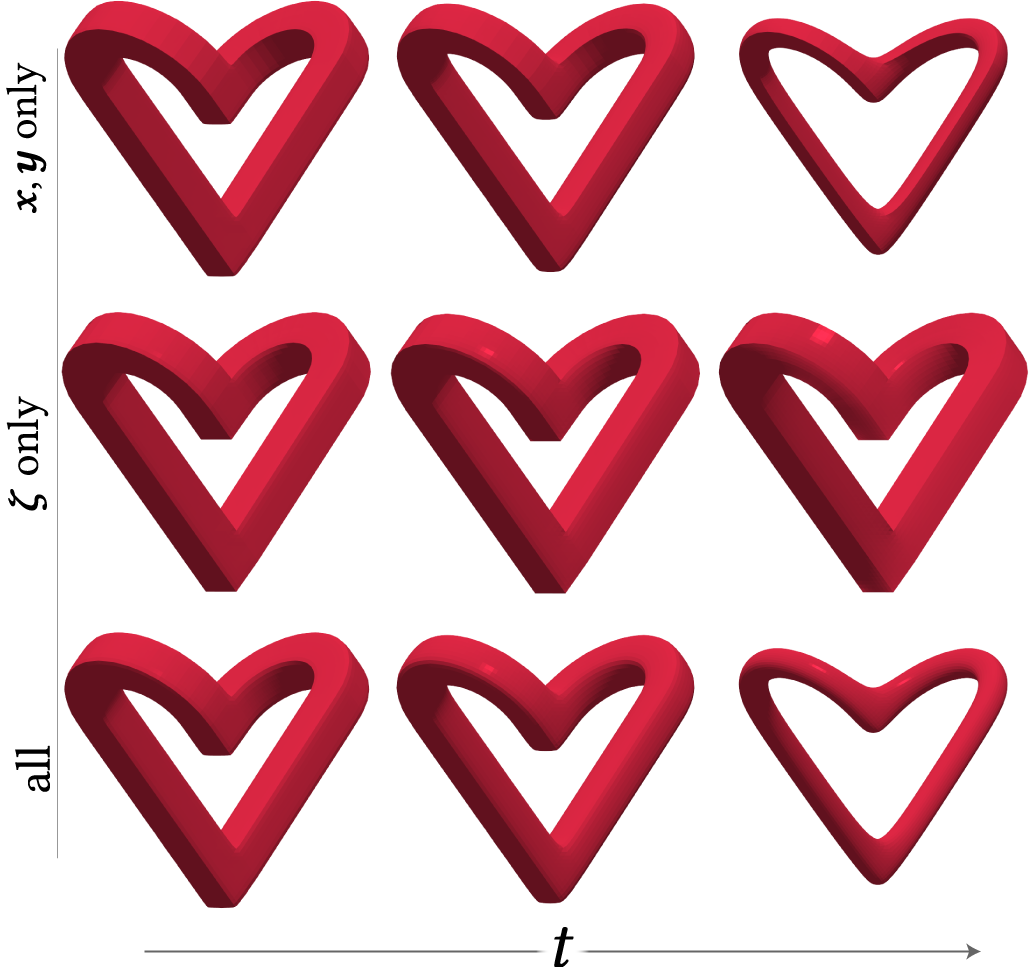}
  \caption{\textsc{Heart.} We can apply Laplacian smoothing to meschers (Section~\ref{ss:smooth}). Smoothing can be used for stylization, or to smooth out modeling imperfections. Here, we subdivide and smooth the mescher \textsc{Heart} using various values of the heat diffusion parameter $t$. We can independently smooth the screen-space components (top row) and depth component (middle row), or smooth both jointly (bottom row).}
  \Description{Heart.}\label{fig:heart}
\end{figure}

To complete our smoothing operator, it is helpful to leverage the structure of $\vect{\zeta}$.
Recall from Equation~\ref{eq:hodge} and Section~\ref{ss:obtaining} that $\vect{\zeta}$ consists of a harmonic component $\vect{\omega}$, and a curl-free component, $\Diff_{01} \vect{z}$.
We know that the harmonic component cannot be further smoothed in this manner, as it is in the null space of the $k$-form Laplacian.
This conveniently leaves only the curl-free component, $\Diff_{01} \vect{z}$, which can be obtained via a Hodge decomposition of $\zeta$.
Since $\vect{z}$ is a primal $0$-form, we can apply standard Laplacian smoothing to it and then combine it back with $\omega$.
In summary, our smoothing operator for $\vect{\zeta}$ is,
\begin{equation}
    \vect{\zeta}_1 = \vect{\omega}_0 + \Diff_{0, 1} \left((I + t\Laplacian)^{-1} \vect{z} - \vect{z}\right).
\end{equation}
Since meschers lack a divergence-free component, this species of smoothing is equivalent to smoothing using the $1$-form Hodge Laplacian in Section~\ref{sec:localintegrability}.
We demonstrate the effect of our smoothing operator with varying degrees of smoothing in Figure~\ref{fig:heart}.

\subsection{Inverse Rendering}\label{ss:inverse}

Robustly inverse-rendering even standard (embeddable) triangle meshes is an active area of research: the key challenge is that gradients are sparse, only defined on the silhouette of an object \citep{li2018differentiable}. One way to make inverse rendering more robust is to use Sobolev gradient descent \cite{Nicolet2021Large}, which effectively smooths out the sparse gradients over the surface. This algorithm only requires a Laplacian---hence, it can be applied to meschers.

Given a ground truth RGB image, we optimize a mescher's geometry so that it matches the target image when passed through the SoftRas differentiable rasterizer \cite{liu2019softras} with fixed lighting. 
Specifically, we optimize the mean squared error between the rendered image and the target, and use separate $\sigma$ parameters in SoftRas for optimizing the screen-space $(x, y)$ coordinates and the relative depth $\zeta$.
In practice, this means that alternate between descending on $\vect{x}, \vect{y}$, and on $\vect{\zeta}$, rendering images using different $\sigma$ for each.
Following \citet{Nicolet2021Large}, we optimize this energy using Sobolev gradients.
Assuming $\vect{g}$ is the gradient with respect to $\vect{x}$ and $\vect{y}$ of our energy, we compute the Sobolev gradient $\hat{\vect{g}}$ as:
\begin{equation}
    \hat{\vect{g}} = (I + \lambda\Laplacian)^{-1} \vect{g}.
\end{equation}
We use the same diffusion parameter $\lambda$ as \citet{Nicolet2021Large}.

During inverse rendering, we discard the partial ordering graph data structure and rely on a differentiable global ordering by assigning a scalar depth offset to each triangle. 
A partial ordering can still be extracted after the optimization by searching for overlaps and comparing triangle scalar depth values.

\begin{figure}[ht]
  \centering
  \includegraphics[width=\linewidth]{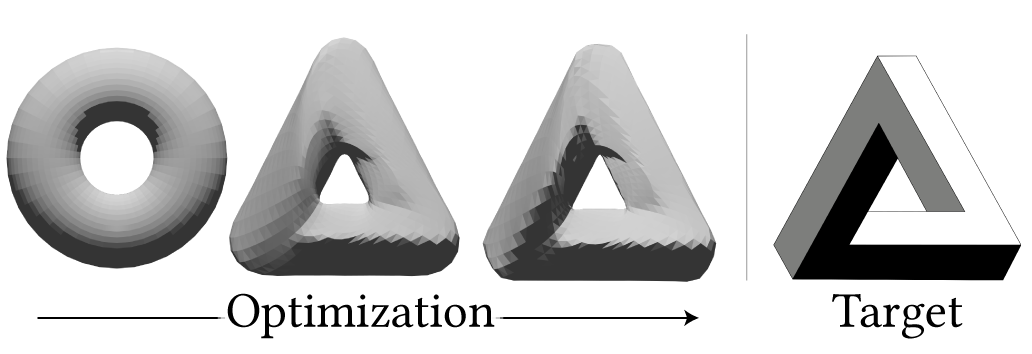}
  \caption{\textsc{Penrose triangle.} We can use gradient-based inverse rendering to recover the Penrose triangle from a source image (Section~\ref{ss:inverse}). The resulting mescher really does represent impossible geometry: that is, we can verify that it has a nonzero harmonic component $\vect{\omega}$.}
  \Description{Inverse rendering.}\label{fig:inverse}
\end{figure}

Figure~\ref{fig:inverse} demonstrates a proof-of-concept of our inverse rendering pipeline.
We initialize our shape to a standard (i.e., possible) torus which is then optimized to look more like the impossible triangle.
Finally, we can check that we are indeed recovering an impossible object by examining the mescher's harmonic component.

\subsection{Embedding Meschers}\label{sec:embedding}

We can easily convert meschers into the two existing representations of impossible objects discussed in Section~\ref{sec:prior-work-rendering}.
\begin{itemize}
  \item To recover a \emph{cut} embedding, we can allow a user to specify a cut on the mescher. Then, we integrate the surface around that cut using a simple breadth-first traversal on its edges.
  \item To recover a \emph{bent} embedding of a mescher, we have to remove the harmonic component from $\vect{\zeta}$ and then globally integrate the resulting field. Luckily, that integral is already given to us as part of the Hodge decomposition in the form of $\vect{z}$ from Equation~\ref{eq:hodge}. Therefore, $\vect{z}$ is a valid choice for the depth component of a mescher.
\end{itemize}
In this way, meschers are a natural generalization of both of these approaches. In fact, they capture the ``best of all worlds'': as we show in Figure~\ref{fig:comp}, meschers are the only representation that can support the full set of rendering and geometry processing operations we consider. For example, while all three representations support rendering, the bent representation leads to artifacts when relighting, and both cut and bent representations lead to artifacts when smoothing and when computing geodesic distances.

\section{Implementation Details}\label{sec:implementation}

We implement Meschers primarily using PyTorch~\cite{pytorch} and relied on the NetworkX~\cite{hagberg2008networkx} graph processing library for the depth ordering.
For all of our experiments, we used a computer with an Intel i9-13900 CPU, $32$ GB of memory, and an Nvidia GeForce RTX 4090.
To implement a user interface and rendering capabilities, we relied on the ModernGL \citep{Dombi2020} and Dear ImGUI \citep{dearimgui} libraries.
We use PyTorch3D \cite{ravi2020pytorch3d} for its implementation of Softras \cite{liu2019softras}.

\section{Discussion, Limitations, and future work}\label{sec:discussion}

In this paper, we presented a new geometric representation for impossible objects. Our representation, the mescher, is designed around the insight that human visual perception is local rather than global. The mescher thus represents locally-consistent geometry while relaxing the requirement that the geometry be globally-consistent. Conveniently, local consistency is all that is needed for a wide variety of classic geometry processing algorithms. Hence, the mescher allows for not only rendering and relighting, but also operations like subdivision, Laplacian smoothing, heat diffusion, geodesic distance queries, and inverse rendering, in a way that aligns with our perceptual intuitions.

\setlength{\columnsep}{4pt}%
\begin{wrapfigure}{r}{0.31\linewidth}
\begin{center}
\includegraphics[width=\linewidth]{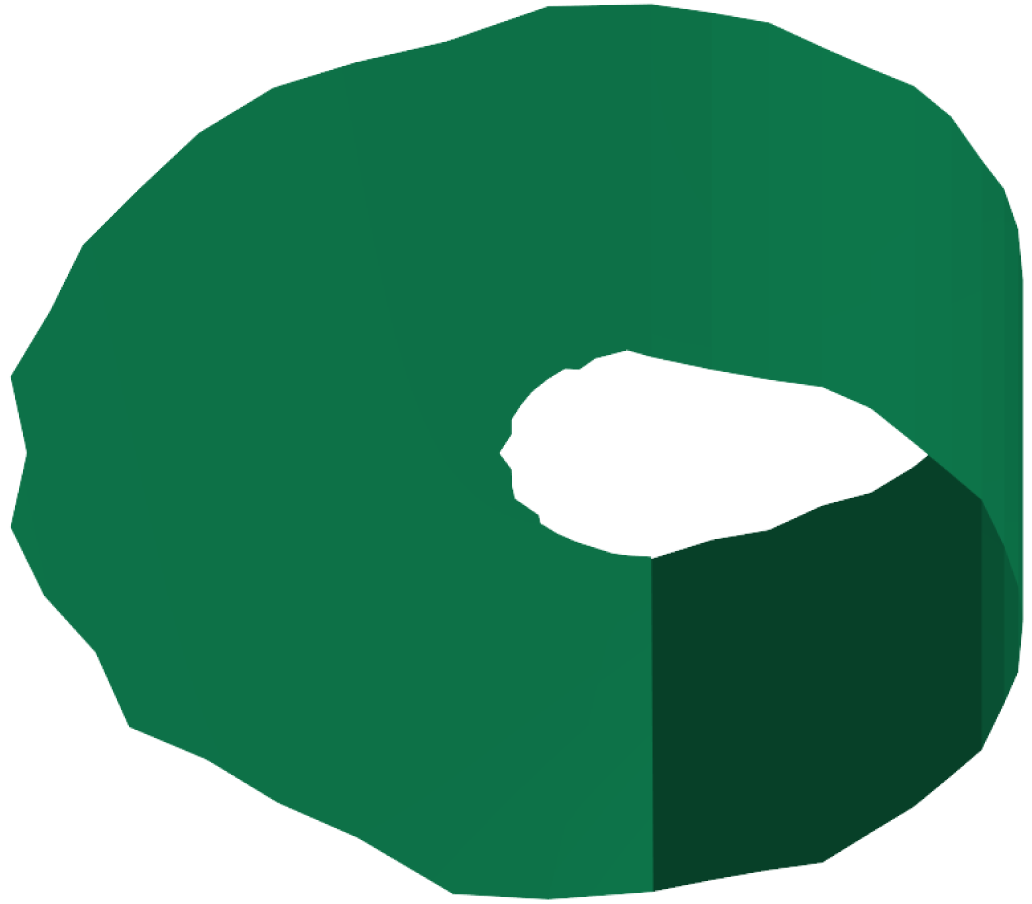}
\end{center}
\end{wrapfigure}
To achieve this, we relied on DEC, which necessitates that its inputs are orientable and manifold.
If we wished to represent a non-orientable impossible object, e.g., if we wished to insert the \textsc{Impossible Lettuce} (see inset) into the bagel in Figure~\ref{fig:teaser}, we would have to consider alternative formulations.

All of our meschers were modeled in an external tool, and then made impossible using our software (Section~\ref{ss:obtaining}).
It remains an open question of what a usable interface for directly modeling meschers is.
An alternative to modeling meschers is to inverse render them, as discussed in Section~\ref{ss:inverse}.
While our experiment serves as the, to our knowledge, first proof of concept of rendering a mescher, we trust that there is 
still room for improvement on this front.
An interesting avenue for future work would be to create a mescher modeling interface that relies on inverse rendering, by, e.g., allowing users to paint a harmonic component onto a possible object.

There are many natural steps for improving meschers' renderings. 
For example, have not experimented with rotating meschers as when we apply a rotation to the differential coordinates, the harmonic component becomes ``mixed in'' into $\vect{x}$ and $\vect{y}$, which would require reprojecting them back onto the space of exact forms.
Furthermore, one can imagine combining meschers with previous work to render them with shadows or transparency.

Beyond rendering, our work has only scratched the surface of geometry processing applications enabled by the mescher representation.
While we compute gradients on a mescher in Section~\ref{ss:heat}, there might be other extrinsic operations one might want to perform, such as, e.g., geometric flows.
Similarly, DEC is a first-order discretization of a surface and there might be applications, such as computing occluding contours of a mescher \cite{capouellez2023contours}, where a higher order discretization might be useful.

In computer vision, implicit geometry representations (e.g.\ signed distance functions) have seen great success due to their ability to change topology during optimization, posing the natural question of if it is possible to develop a mescher-like implicit representation.

Lastly, since our representation is inspired by vision science, we can conversely ask whether it would be in turn useful to vision science research.
For example, an exciting question one might ask is whether shortest paths computed on a mescher match those that a human would manually trace out.

Perceptually-informed and human-centered models can aid the understanding human vision and the design of computer vision systems, and are central in creativity and art.
Through our study of impossible objects, we hope to encourage work not only on realistic graphics models, but also those that become possible in the context of human perception.

\begin{acks}
We would like to thank Tzu-Mao Li, Jeffrey Price and Seth Riskin for thoughtful conversations during this project. 

Ana Dodik acknowledges the generous support of the MIT Presidential Fellowship and the Mathworks Fellowship. Kartik Chandra acknowledges the Hertz Foundation and the NSF GRFP for funding.

Jonathan Ragan-Kelley acknowledges the support of NSF award OAC-2403239.

Joshua Tenenbaum acknowledges the support of Schmidt Sciences AI2050 fellowship and Siegel Family Quest for Intelligence.

The MIT Geometric Data Processing group acknowledges the generous support of Army Research Office grants W911NF2010168 and W911NF2110293, of Air Force Office of Scientific Research award FA9550-19-1-031, of National Science Foundation grant CHS-1955697, from the CSAIL Systems that Learn program, from the MIT–IBM Watson AI Laboratory, from the Toyota–CSAIL Joint Research Center, from a gift from Adobe Systems, and from a Google Research Scholar award.

The MIT Scene Representation Group was supported by the National Science Foundation under Grant No. 2211259, by the Singapore DSTA under DST00OECI20300823 (New Representations for Vision, 3D Self-Supervised Learning for Label-Efficient Vision), by the Intelligence Advanced Research Projects Activity (IARPA) via Department of Interior/ Interior Business Center (DOI/IBC) under 140D0423C0075, by the Amazon Science Hub, and by the MIT-Google Program for Computing Innovation.
\end{acks}

\bibliographystyle{ACM-Reference-Format}
\bibliography{bibliography}


\begin{thebibliography}{75}

%%% the Web address.

\ifx \showCODEN    \undefined \def \showCODEN     #1{\unskip}     \fi
\ifx \showDOI      \undefined \def \showDOI       #1{#1}\fi
\ifx \showISBNx    \undefined \def \showISBNx     #1{\unskip}     \fi
\ifx \showISBNxiii \undefined \def \showISBNxiii  #1{\unskip}     \fi
\ifx \showISSN     \undefined \def \showISSN      #1{\unskip}     \fi
\ifx \showLCCN     \undefined \def \showLCCN      #1{\unskip}     \fi
\ifx \shownote     \undefined \def \shownote      #1{#1}          \fi
\ifx \showarticletitle \undefined \def \showarticletitle #1{#1}   \fi
\ifx \showURL      \undefined \def \showURL       {\relax}        \fi
% invisible to TeX
\providecommand\bibfield[2]{#2}
\providecommand\bibinfo[2]{#2}
\providecommand\natexlab[1]{#1}
\providecommand\showeprint[2][]{arXiv:#2}

\bibitem[Agarwala et~al\mbox{.}(2006)]%
        {agarwala2006photographing}
\bibfield{author}{\bibinfo{person}{Aseem Agarwala}, \bibinfo{person}{Maneesh
  Agrawala}, \bibinfo{person}{Michael Cohen}, \bibinfo{person}{David Salesin},
  {and} \bibinfo{person}{Richard Szeliski}.} \bibinfo{year}{2006}\natexlab{}.
\newblock \showarticletitle{Photographing long scenes with multi-viewpoint
  panoramas}.
\newblock In \bibinfo{booktitle}{\emph{ACM SIGGRAPH 2006 Papers}}.
  \bibinfo{pages}{853--861}.
\newblock


\bibitem[Agrawala et~al\mbox{.}(2000)]%
        {agrawala2000artistic}
\bibfield{author}{\bibinfo{person}{Maneesh Agrawala}, \bibinfo{person}{Denis
  Zorin}, {and} \bibinfo{person}{Tamara Munzner}.}
  \bibinfo{year}{2000}\natexlab{}.
\newblock \showarticletitle{Artistic multiprojection rendering}. In
  \bibinfo{booktitle}{\emph{Rendering Techniques 2000: Proceedings of the
  Eurographics Workshop in Brno, Czech Republic, June 26--28, 2000 11}}.
  Springer, \bibinfo{pages}{125--136}.
\newblock


\bibitem[Alexa(2003)]%
        {alexa2003differential}
\bibfield{author}{\bibinfo{person}{Marc Alexa}.}
  \bibinfo{year}{2003}\natexlab{}.
\newblock \showarticletitle{Differential coordinates for local mesh morphing
  and deformation}.
\newblock \bibinfo{journal}{\emph{The Visual Computer}} \bibinfo{volume}{19},
  \bibinfo{number}{2} (\bibinfo{year}{2003}), \bibinfo{pages}{105--114}.
\newblock


\bibitem[Alexeev(2001)]%
        {impossibleImpossibleWorld}
\bibfield{author}{\bibinfo{person}{Vlad Alexeev}.}
  \bibinfo{year}{2001}\natexlab{}.
\newblock \bibinfo{booktitle}{\emph{{I}mpossible world --- im-possible.info}}.
\newblock
\newblock
\shownote{[Accessed 16-04-2025]}.


\bibitem[Biederman(1985)]%
        {biederman1985human}
\bibfield{author}{\bibinfo{person}{Irving Biederman}.}
  \bibinfo{year}{1985}\natexlab{}.
\newblock \showarticletitle{Human image understanding: Recent research and a
  theory}.
\newblock \bibinfo{journal}{\emph{Computer vision, graphics, and image
  processing}} \bibinfo{volume}{32}, \bibinfo{number}{1}
  (\bibinfo{year}{1985}), \bibinfo{pages}{29--73}.
\newblock


\bibitem[Bommes et~al\mbox{.}(2023)]%
        {bommes2023mixed}
\bibfield{author}{\bibinfo{person}{David Bommes}, \bibinfo{person}{Henrik
  Zimmer}, {and} \bibinfo{person}{Leif Kobbelt}.}
  \bibinfo{year}{2023}\natexlab{}.
\newblock \bibinfo{booktitle}{\emph{Mixed-Integer Quadrangulation}
  (\bibinfo{edition}{1} ed.)}.
\newblock \bibinfo{publisher}{Association for Computing Machinery},
  \bibinfo{address}{New York, NY, USA}.
\newblock
\showISBNx{9798400708978}
\urldef\tempurl%
\url{https://doi.org/10.1145/3596711.3596740}
\showURL{%
\tempurl}


\bibitem[Burgert et~al\mbox{.}(2023)]%
        {burgert2023diffusion}
\bibfield{author}{\bibinfo{person}{Ryan Burgert}, \bibinfo{person}{Xiang Li},
  \bibinfo{person}{Abe Leite}, \bibinfo{person}{Kanchana Ranasinghe}, {and}
  \bibinfo{person}{Michael~S Ryoo}.} \bibinfo{year}{2023}\natexlab{}.
\newblock \showarticletitle{Diffusion illusions: Hiding images in plain sight}.
\newblock \bibinfo{journal}{\emph{arXiv preprint arXiv:2312.03817}}
  (\bibinfo{year}{2023}).
\newblock


\bibitem[Capouellez et~al\mbox{.}(2023)]%
        {capouellez2023contours}
\bibfield{author}{\bibinfo{person}{Ryan Capouellez}, \bibinfo{person}{Jiacheng
  Dai}, \bibinfo{person}{Aaron Hertzmann}, {and} \bibinfo{person}{Denis
  Zorin}.} \bibinfo{year}{2023}\natexlab{}.
\newblock \showarticletitle{Algebraic Smooth Occluding Contours}. In
  \bibinfo{booktitle}{\emph{ACM SIGGRAPH 2023 Conference Proceedings}} (Los
  Angeles, CA, USA) \emph{(\bibinfo{series}{Siggraph '23})}.
  \bibinfo{publisher}{Association for Computing Machinery},
  \bibinfo{address}{New York, NY, USA}, Article \bibinfo{articleno}{39},
  \bibinfo{numpages}{10}~pages.
\newblock
\showISBNx{9798400701597}
\urldef\tempurl%
\url{https://doi.org/10.1145/3588432.3591547}
\showDOI{\tempurl}


\bibitem[Cavanagh(2005)]%
        {cavanagh2005artist}
\bibfield{author}{\bibinfo{person}{Patrick Cavanagh}.}
  \bibinfo{year}{2005}\natexlab{}.
\newblock \showarticletitle{The artist as neuroscientist}.
\newblock \bibinfo{journal}{\emph{Nature}} \bibinfo{volume}{434},
  \bibinfo{number}{7031} (\bibinfo{year}{2005}), \bibinfo{pages}{301--307}.
\newblock


\bibitem[Chandra et~al\mbox{.}(2022)]%
        {chandra2022designing}
\bibfield{author}{\bibinfo{person}{Kartik Chandra}, \bibinfo{person}{Tzu-Mao
  Li}, \bibinfo{person}{Joshua Tenenbaum}, {and} \bibinfo{person}{Jonathan
  Ragan-Kelley}.} \bibinfo{year}{2022}\natexlab{}.
\newblock \showarticletitle{Designing perceptual puzzles by differentiating
  probabilistic programs}. In \bibinfo{booktitle}{\emph{ACM SIGGRAPH 2022
  Conference Proceedings}}. \bibinfo{pages}{1--9}.
\newblock


\bibitem[Chi et~al\mbox{.}(2008)]%
        {chi2008self}
\bibfield{author}{\bibinfo{person}{Ming-Te Chi}, \bibinfo{person}{Tong-Yee
  Lee}, \bibinfo{person}{Yingge Qu}, {and} \bibinfo{person}{Tien-Tsin Wong}.}
  \bibinfo{year}{2008}\natexlab{}.
\newblock \showarticletitle{Self-Animating Images: Illusory Motion Using
  Repeated Asymmetric Patterns}.
\newblock \bibinfo{journal}{\emph{ACM Trans. Graph.}} \bibinfo{volume}{27},
  \bibinfo{number}{3} (\bibinfo{date}{aug} \bibinfo{year}{2008}),
  \bibinfo{pages}{1–8}.
\newblock
\showISSN{0730-0301}
\urldef\tempurl%
\url{https://doi.org/10.1145/1360612.1360661}
\showDOI{\tempurl}


\bibitem[Chu et~al\mbox{.}(2010)]%
        {chu2010camouflage}
\bibfield{author}{\bibinfo{person}{Hung-Kuo Chu}, \bibinfo{person}{Wei-Hsin
  Hsu}, \bibinfo{person}{Niloy~J Mitra}, \bibinfo{person}{Daniel Cohen-Or},
  \bibinfo{person}{Tien-Tsin Wong}, {and} \bibinfo{person}{Tong-Yee Lee}.}
  \bibinfo{year}{2010}\natexlab{}.
\newblock \showarticletitle{Camouflage images.}
\newblock \bibinfo{journal}{\emph{ACM Trans. Graph.}} \bibinfo{volume}{29},
  \bibinfo{number}{4} (\bibinfo{year}{2010}), \bibinfo{pages}{51--1}.
\newblock
\urldef\tempurl%
\url{https://dl.acm.org/doi/abs/10.1145/1833349.1778788}
\showURL{%
\tempurl}


\bibitem[Cornut(2014)]%
        {dearimgui}
\bibfield{author}{\bibinfo{person}{Omar Cornut}.}
  \bibinfo{year}{2014}\natexlab{}.
\newblock \bibinfo{howpublished}{https://github.com/ocornut/imgui}.
\newblock


\bibitem[Crane et~al\mbox{.}(2013)]%
        {crane2013deccourse}
\bibfield{author}{\bibinfo{person}{Keenan Crane}, \bibinfo{person}{Fernando de
  Goes}, \bibinfo{person}{Mathieu Desbrun}, {and} \bibinfo{person}{Peter
  Schr\"{o}der}.} \bibinfo{year}{2013}\natexlab{}.
\newblock \showarticletitle{Digital geometry processing with discrete exterior
  calculus}. In \bibinfo{booktitle}{\emph{ACM SIGGRAPH 2013 Courses}} (Anaheim,
  California) \emph{(\bibinfo{series}{Siggraph '13})}.
  \bibinfo{publisher}{Association for Computing Machinery},
  \bibinfo{address}{New York, NY, USA}, Article \bibinfo{articleno}{7},
  \bibinfo{numpages}{126}~pages.
\newblock
\showISBNx{9781450323390}
\urldef\tempurl%
\url{https://doi.org/10.1145/2504435.2504442}
\showDOI{\tempurl}


\bibitem[Crane et~al\mbox{.}(2017)]%
        {Crane:2017:HMD}
\bibfield{author}{\bibinfo{person}{Keenan Crane}, \bibinfo{person}{Clarisse
  Weischedel}, {and} \bibinfo{person}{Max Wardetzky}.}
  \bibinfo{year}{2017}\natexlab{}.
\newblock \showarticletitle{The Heat Method for Distance Computation}.
\newblock \bibinfo{journal}{\emph{Commun. ACM}} \bibinfo{volume}{60},
  \bibinfo{number}{11} (\bibinfo{date}{Oct.} \bibinfo{year}{2017}),
  \bibinfo{pages}{90--99}.
\newblock
\showISSN{0001-0782}
\urldef\tempurl%
\url{https://doi.org/10.1145/3131280}
\showDOI{\tempurl}


\bibitem[de~Goes et~al\mbox{.}(2016)]%
        {deGoes2016subdivision}
\bibfield{author}{\bibinfo{person}{Fernando de Goes}, \bibinfo{person}{Mathieu
  Desbrun}, \bibinfo{person}{Mark Meyer}, {and} \bibinfo{person}{Tony DeRose}.}
  \bibinfo{year}{2016}\natexlab{}.
\newblock \showarticletitle{Subdivision exterior calculus for geometry
  processing}.
\newblock \bibinfo{journal}{\emph{ACM Trans. Graph.}} \bibinfo{volume}{35},
  \bibinfo{number}{4}, Article \bibinfo{articleno}{133} (\bibinfo{date}{July}
  \bibinfo{year}{2016}), \bibinfo{numpages}{11}~pages.
\newblock
\showISSN{0730-0301}
\urldef\tempurl%
\url{https://doi.org/10.1145/2897824.2925880}
\showDOI{\tempurl}


\bibitem[Desbrun et~al\mbox{.}(2006)]%
        {desbrun2006dec}
\bibfield{author}{\bibinfo{person}{Mathieu Desbrun}, \bibinfo{person}{Eva
  Kanso}, {and} \bibinfo{person}{Yiying Tong}.}
  \bibinfo{year}{2006}\natexlab{}.
\newblock \showarticletitle{Discrete differential forms for computational
  modeling}. In \bibinfo{booktitle}{\emph{ACM SIGGRAPH 2006 Courses}} (Boston,
  Massachusetts) \emph{(\bibinfo{series}{Siggraph '06})}.
  \bibinfo{publisher}{Association for Computing Machinery},
  \bibinfo{address}{New York, NY, USA}, \bibinfo{pages}{39–54}.
\newblock
\showISBNx{1595933646}
\urldef\tempurl%
\url{https://doi.org/10.1145/1185657.1185665}
\showDOI{\tempurl}


\bibitem[Desbrun et~al\mbox{.}(2005)]%
        {desbrun2005poincare}
\bibfield{author}{\bibinfo{person}{Mathieu Desbrun}, \bibinfo{person}{Melvin
  Leok}, {and} \bibinfo{person}{Jerrold~E. Marsden}.}
  \bibinfo{year}{2005}\natexlab{}.
\newblock \showarticletitle{Discrete Poincaré Lemma}.
\newblock \bibinfo{journal}{\emph{Applied Numerical Mathematics}}
  \bibinfo{volume}{53}, \bibinfo{number}{2-4} (\bibinfo{date}{May}
  \bibinfo{year}{2005}), \bibinfo{pages}{231–248}.
\newblock
\showISSN{0168-9274}
\urldef\tempurl%
\url{https://doi.org/10.1016/j.apnum.2004.09.035}
\showDOI{\tempurl}


\bibitem[Desbrun et~al\mbox{.}(2023)]%
        {desbrun2023fairing}
\bibfield{author}{\bibinfo{person}{Mathieu Desbrun}, \bibinfo{person}{Mark
  Meyer}, \bibinfo{person}{Peter Schroder}, {and} \bibinfo{person}{Alan~H.
  Barr}.} \bibinfo{year}{2023}\natexlab{}.
\newblock \bibinfo{booktitle}{\emph{Implicit Fairing of Irregular Meshes using
  Diffusion and Curvature Flow} (\bibinfo{edition}{1} ed.)}.
\newblock \bibinfo{publisher}{Association for Computing Machinery},
  \bibinfo{address}{New York, NY, USA}.
\newblock
\showISBNx{9798400708978}
\urldef\tempurl%
\url{https://doi.org/10.1145/3596711.3596729}
\showURL{%
\tempurl}


\bibitem[Dombi(2020)]%
        {Dombi2020}
\bibfield{author}{\bibinfo{person}{Szabolcs Dombi}.}
  \bibinfo{year}{2020}\natexlab{}.
\newblock \bibinfo{title}{ModernGL, high performance python bindings for OpenGL
  3.3+}.
\newblock \bibinfo{howpublished}{\url{https://github.com/moderngl/moderngl}}.
\newblock


\bibitem[Draper(1978)]%
        {Draper1978Family}
\bibfield{author}{\bibinfo{person}{Stephen~W Draper}.}
  \bibinfo{year}{1978}\natexlab{}.
\newblock \showarticletitle{The Penrose Triangle and a Family of Related
  Figures}.
\newblock \bibinfo{journal}{\emph{Perception}} \bibinfo{volume}{7},
  \bibinfo{number}{3} (\bibinfo{year}{1978}), \bibinfo{pages}{283--296}.
\newblock
\urldef\tempurl%
\url{https://doi.org/10.1068/p070283}
\showDOI{\tempurl}
\showeprint{https://doi.org/10.1068/p070283}
\newblock
\shownote{Pmid: 693228}.


\bibitem[Elber(2011)]%
        {elber2011modeling}
\bibfield{author}{\bibinfo{person}{Gershon Elber}.}
  \bibinfo{year}{2011}\natexlab{}.
\newblock \showarticletitle{Modeling (seemingly) impossible models}.
\newblock \bibinfo{journal}{\emph{Computers \& Graphics}} \bibinfo{volume}{35},
  \bibinfo{number}{3} (\bibinfo{year}{2011}), \bibinfo{pages}{632--638}.
\newblock


\bibitem[Elber(2012)]%
        {gershonelberEscherReal}
\bibfield{author}{\bibinfo{person}{Gershon Elber}.}
  \bibinfo{year}{2012}\natexlab{}.
\newblock \bibinfo{booktitle}{\emph{{E}scher for {R}eal --- gershonelber.org}}.
\newblock
\newblock
\shownote{[Accessed 16-04-2025]}.


\bibitem[Ernst(1986)]%
        {ernst1986adventures}
\bibfield{author}{\bibinfo{person}{B. Ernst}.} \bibinfo{year}{1986}\natexlab{}.
\newblock \bibinfo{booktitle}{\emph{Adventures with Impossible Figures}}.
\newblock \bibinfo{publisher}{Tarquin}.
\newblock
\showISBNx{9780906212547}
\urldef\tempurl%
\url{https://books.google.com/books?id=HU_vAAAAMAAJ}
\showURL{%
\tempurl}


\bibitem[Escher(1961)]%
        {escher1961graphic}
\bibfield{author}{\bibinfo{person}{M.C. Escher}.}
  \bibinfo{year}{1961}\natexlab{}.
\newblock \bibinfo{booktitle}{\emph{The Graphic Work of M.C. Escher}}.
\newblock \bibinfo{publisher}{Oldbourne}.
\newblock
\showLCCN{gb61014941}
\urldef\tempurl%
\url{https://books.google.com/books?id=1T5gwgEACAAJ}
\showURL{%
\tempurl}


\bibitem[Freeman et~al\mbox{.}(1991)]%
        {freeman1991motion}
\bibfield{author}{\bibinfo{person}{William~T Freeman},
  \bibinfo{person}{Edward~H Adelson}, {and} \bibinfo{person}{David~J Heeger}.}
  \bibinfo{year}{1991}\natexlab{}.
\newblock \showarticletitle{Motion without movement}.
\newblock \bibinfo{journal}{\emph{ACM Siggraph Computer Graphics}}
  \bibinfo{volume}{25}, \bibinfo{number}{4} (\bibinfo{year}{1991}),
  \bibinfo{pages}{27--30}.
\newblock


\bibitem[Freud et~al\mbox{.}(2013)]%
        {freud2013b}
\bibfield{author}{\bibinfo{person}{Erez Freud}, \bibinfo{person}{Tzvi Ganel},
  {and} \bibinfo{person}{Galia Avidan}.} \bibinfo{year}{2013}\natexlab{}.
\newblock \showarticletitle{Representation of possible and impossible objects
  in the human visual cortex: Evidence from fMRI adaptation}.
\newblock \bibinfo{journal}{\emph{NeuroImage}}  \bibinfo{volume}{64}
  (\bibinfo{year}{2013}), \bibinfo{pages}{685--692}.
\newblock
\showISSN{1053-8119}
\urldef\tempurl%
\url{https://doi.org/10.1016/j.neuroimage.2012.08.070}
\showDOI{\tempurl}


\bibitem[Freud et~al\mbox{.}(2015)]%
        {freud2015}
\bibfield{author}{\bibinfo{person}{Erez Freud}, \bibinfo{person}{Bat-Sheva
  Hadad}, \bibinfo{person}{Galia Avidan}, {and} \bibinfo{person}{Tzvi Ganel}.}
  \bibinfo{year}{2015}\natexlab{}.
\newblock \showarticletitle{Evidence for similar early but not late
  representation of possible and impossible objects}.
\newblock \bibinfo{journal}{\emph{Frontiers in Psychology}}
  \bibinfo{volume}{6} (\bibinfo{year}{2015}).
\newblock
\showISSN{1664-1078}
\urldef\tempurl%
\url{https://doi.org/10.3389/fpsyg.2015.00094}
\showDOI{\tempurl}


\bibitem[G\'arate(2020)]%
        {sketchfabSpotlightImpossible}
\bibfield{author}{\bibinfo{person}{Mat\'ias G\'arate}.}
  \bibinfo{year}{2020}\natexlab{}.
\newblock \bibinfo{booktitle}{\emph{{A}rt {S}potlight: {I}mpossible {P}enrose
  {S}nake --- sketchfab.com}}.
\newblock
\newblock
\shownote{[Accessed 16-04-2025]}.


\bibitem[G\'arate(2023)]%
        {blenderOutsModelling}
\bibfield{author}{\bibinfo{person}{Mat\'ias G\'arate}.}
  \bibinfo{year}{2023}\natexlab{}.
\newblock \bibinfo{booktitle}{\emph{{T}he ins and outs of modelling impossible
  figures. — {B}lender {C}onference 2023 --- conference.blender.org}}.
\newblock
\newblock
\shownote{[Accessed 16-04-2025]}.


\bibitem[Geng et~al\mbox{.}(2023)]%
        {geng2023visual}
\bibfield{author}{\bibinfo{person}{Daniel Geng}, \bibinfo{person}{Inbum Park},
  {and} \bibinfo{person}{Andrew Owens}.} \bibinfo{year}{2023}\natexlab{}.
\newblock \showarticletitle{Visual Anagrams: Generating Multi-View Optical
  Illusions with Diffusion Models}.
\newblock \bibinfo{journal}{\emph{arXiv preprint arXiv:2311.17919}}
  (\bibinfo{year}{2023}).
\newblock


\bibitem[Geng et~al\mbox{.}(2024)]%
        {geng2024factorized}
\bibfield{author}{\bibinfo{person}{Daniel Geng}, \bibinfo{person}{Inbum Park},
  {and} \bibinfo{person}{Andrew Owens}.} \bibinfo{year}{2024}\natexlab{}.
\newblock \showarticletitle{Factorized Diffusion: Perceptual Illusions by Noise
  Decomposition}.
\newblock \bibinfo{journal}{\emph{arXiv preprint arXiv:2404.11615}}
  (\bibinfo{year}{2024}).
\newblock


\bibitem[Gregory(1970)]%
        {Gregory1970Eye}
\bibfield{author}{\bibinfo{person}{R.~L. (Richard~Langton) Gregory}.}
  \bibinfo{year}{1970}\natexlab{}.
\newblock \bibinfo{booktitle}{\emph{The intelligent eye, by R. L. Gregory.}}
\newblock \bibinfo{publisher}{Weidenfeld \& Nicolson},
  \bibinfo{address}{London}.
\newblock
\showISBNx{0297000217}
\showLCCN{70484481}


\bibitem[Hagberg et~al\mbox{.}(2008)]%
        {hagberg2008networkx}
\bibfield{author}{\bibinfo{person}{Aric~A. Hagberg}, \bibinfo{person}{Daniel~A.
  Schult}, {and} \bibinfo{person}{Pieter~J. Swart}.}
  \bibinfo{year}{2008}\natexlab{}.
\newblock \showarticletitle{Exploring Network Structure, Dynamics, and Function
  using NetworkX}. In \bibinfo{booktitle}{\emph{Proceedings of the 7th Python
  in Science Conference}}, \bibfield{editor}{\bibinfo{person}{Ga\"el
  Varoquaux}, \bibinfo{person}{Travis Vaught}, {and} \bibinfo{person}{Jarrod
  Millman}} (Eds.). \bibinfo{address}{Pasadena, CA USA},
  \bibinfo{pages}{11--15}.
\newblock
\urldef\tempurl%
\url{http://conference.scipy.org/proceedings/SciPy2008/paper_2/}
\showURL{%
\tempurl}


\bibitem[Heinke et~al\mbox{.}(2021)]%
        {Heinke2021}
\bibfield{author}{\bibinfo{person}{Dietmar Heinke}, \bibinfo{person}{Peter
  Wachman}, \bibinfo{person}{Wieske {van Zoest}}, {and}
  \bibinfo{person}{E.~Charles Leek}.} \bibinfo{year}{2021}\natexlab{}.
\newblock \showarticletitle{A failure to learn object shape geometry:
  Implications for convolutional neural networks as plausible models of
  biological vision}.
\newblock \bibinfo{journal}{\emph{Vision Research}}  \bibinfo{volume}{189}
  (\bibinfo{year}{2021}), \bibinfo{pages}{81--92}.
\newblock
\showISSN{0042-6989}
\urldef\tempurl%
\url{https://doi.org/10.1016/j.visres.2021.09.004}
\showDOI{\tempurl}


\bibitem[Hertzmann(2024)]%
        {hertzman2024perspective}
\bibfield{author}{\bibinfo{person}{Aaron Hertzmann}.}
  \bibinfo{year}{2024}\natexlab{}.
\newblock \showarticletitle{{Toward a theory of perspective perception in
  pictures}}.
\newblock \bibinfo{journal}{\emph{Journal of Vision}} \bibinfo{volume}{24},
  \bibinfo{number}{4} (\bibinfo{date}{04} \bibinfo{year}{2024}),
  \bibinfo{pages}{23--23}.
\newblock
\showISSN{1534-7362}
\urldef\tempurl%
\url{https://doi.org/10.1167/jov.24.4.23}
\showDOI{\tempurl}
\showeprint{https://arvojournals.org/arvo/content\_public/journal/jov/938670/i1534-7362-24-4-23\_1713961163.65834.pdf}


\bibitem[Hirani(2003)]%
        {Hirani2003}
\bibfield{author}{\bibinfo{person}{Anil~N. Hirani}.}
  \bibinfo{year}{2003}\natexlab{}.
\newblock \emph{\bibinfo{title}{Discrete Exterior Calculus}}.
\newblock \bibinfo{thesistype}{Ph.\,D. Dissertation}.
  \bibinfo{school}{California Institute of Technology}.
\newblock
\urldef\tempurl%
\url{http://resolver.caltech.edu/CaltechETD:etd-05202003-095403}
\showURL{%
\tempurl}


\bibitem[Hogarth(1754)]%
        {hogarth_satire_1754}
\bibfield{author}{\bibinfo{person}{William Hogarth}.}
  \bibinfo{year}{1754}\natexlab{}.
\newblock \bibinfo{title}{Satire on False Perspective}.
\newblock
\newblock
\urldef\tempurl%
\url{https://publicdomainreview.org/collection/william-hogarth-satire-on-false-perspective/}
\showURL{%
\tempurl}
\newblock
\shownote{Frontispiece to Joshua Kirby's "Dr. Brook Taylor's Method of
  Perspective Made Easy"}.


\bibitem[Inglis(2014)]%
        {inglis2014constructing}
\bibfield{author}{\bibinfo{person}{Tiffany Inglis}.}
  \bibinfo{year}{2014}\natexlab{}.
\newblock \showarticletitle{Constructing Drawings of Impossible Figures with
  Axonometric Blocks and Pseudo-3D Manipulations}. In
  \bibinfo{booktitle}{\emph{Proceedings of Bridges 2014: Mathematics, Music,
  Art, Architecture, Culture}}, \bibfield{editor}{\bibinfo{person}{Gary
  Greenfield}, \bibinfo{person}{George Hart}, {and} \bibinfo{person}{Reza
  Sarhangi}} (Eds.). \bibinfo{publisher}{Tessellations Publishing},
  \bibinfo{address}{Phoenix, Arizona}, \bibinfo{pages}{159--166}.
\newblock
\showISBNx{978-1-938664-11-3}
\showISSN{1099-6702}
\urldef\tempurl%
\url{http://archive.bridgesmathart.org/2014/bridges2014-159.html}
\showURL{%
\tempurl}


\bibitem[Karpenko and Hughes(2006)]%
        {karpenko2006smoothsketch}
\bibfield{author}{\bibinfo{person}{Olga~A Karpenko} {and}
  \bibinfo{person}{John~F Hughes}.} \bibinfo{year}{2006}\natexlab{}.
\newblock \showarticletitle{Smoothsketch: 3d free-form shapes from complex
  sketches}.
\newblock In \bibinfo{booktitle}{\emph{ACM SIGGRAPH 2006 Papers}}.
  \bibinfo{pages}{589--598}.
\newblock


\bibitem[Khoh and Kovesi(1999)]%
        {khoh1999animating}
\bibfield{author}{\bibinfo{person}{Chih~W Khoh} {and} \bibinfo{person}{Peter
  Kovesi}.} \bibinfo{year}{1999}\natexlab{}.
\newblock \bibinfo{title}{Animating impossible objects}.
\newblock
\newblock


\bibitem[Kn\"{o}ppel et~al\mbox{.}(2015)]%
        {Knoppel2015SPS}
\bibfield{author}{\bibinfo{person}{Felix Kn\"{o}ppel}, \bibinfo{person}{Keenan
  Crane}, \bibinfo{person}{Ulrich Pinkall}, {and} \bibinfo{person}{Peter
  Schr\"{o}der}.} \bibinfo{year}{2015}\natexlab{}.
\newblock \showarticletitle{Stripe Patterns on Surfaces}.
\newblock \bibinfo{journal}{\emph{ACM Trans. Graph.}}  \bibinfo{volume}{34}
  (\bibinfo{year}{2015}).
\newblock
Issue 4.


\bibitem[Koenderink(1998)]%
        {koenderink1998pictorial}
\bibfield{author}{\bibinfo{person}{Jan~J Koenderink}.}
  \bibinfo{year}{1998}\natexlab{}.
\newblock \showarticletitle{Pictorial relief}.
\newblock \bibinfo{journal}{\emph{Philosophical Transactions of the Royal
  Society of London. Series A: Mathematical, Physical and Engineering
  Sciences}} \bibinfo{volume}{356}, \bibinfo{number}{1740}
  (\bibinfo{year}{1998}), \bibinfo{pages}{1071--1086}.
\newblock


\bibitem[Kälberer et~al\mbox{.}(2007)]%
        {kaelberer2007quadcover}
\bibfield{author}{\bibinfo{person}{Felix Kälberer}, \bibinfo{person}{Matthias
  Nieser}, {and} \bibinfo{person}{Konrad Polthier}.}
  \bibinfo{year}{2007}\natexlab{}.
\newblock \showarticletitle{QuadCover - Surface Parameterization using Branched
  Coverings}.
\newblock \bibinfo{journal}{\emph{Computer Graphics Forum}}
  \bibinfo{volume}{26}, \bibinfo{number}{3} (\bibinfo{year}{2007}),
  \bibinfo{pages}{375--384}.
\newblock
\urldef\tempurl%
\url{https://doi.org/10.1111/j.1467-8659.2007.01060.x}
\showDOI{\tempurl}
\showeprint{https://onlinelibrary.wiley.com/doi/pdf/10.1111/j.1467-8659.2007.01060.x}


\bibitem[Lai et~al\mbox{.}(2015)]%
        {lai20153d}
\bibfield{author}{\bibinfo{person}{Chi-Fu~William Lai},
  \bibinfo{person}{Sai-Kit Yeung}, \bibinfo{person}{Xiaoqi Yan},
  \bibinfo{person}{Chi-Wing Fu}, {and} \bibinfo{person}{Chi-Keung Tang}.}
  \bibinfo{year}{2015}\natexlab{}.
\newblock \showarticletitle{3D navigation on impossible figures via dynamically
  reconfigurable maze}.
\newblock \bibinfo{journal}{\emph{IEEE transactions on visualization and
  computer graphics}} \bibinfo{volume}{22}, \bibinfo{number}{10}
  (\bibinfo{year}{2015}), \bibinfo{pages}{2275--2288}.
\newblock


\bibitem[Li et~al\mbox{.}(2018)]%
        {li2018differentiable}
\bibfield{author}{\bibinfo{person}{Tzu-Mao Li}, \bibinfo{person}{Miika
  Aittala}, \bibinfo{person}{Fr{\'e}do Durand}, {and} \bibinfo{person}{Jaakko
  Lehtinen}.} \bibinfo{year}{2018}\natexlab{}.
\newblock \showarticletitle{Differentiable monte carlo ray tracing through edge
  sampling}.
\newblock \bibinfo{journal}{\emph{ACM Transactions on Graphics (TOG)}}
  \bibinfo{volume}{37}, \bibinfo{number}{6} (\bibinfo{year}{2018}),
  \bibinfo{pages}{1--11}.
\newblock


\bibitem[Li et~al\mbox{.}(2024)]%
        {li2024possibleimpossibles}
\bibfield{author}{\bibinfo{person}{Yuanbo Li}, \bibinfo{person}{Tianyi Ma},
  \bibinfo{person}{Zaineb Aljumayaat}, {and} \bibinfo{person}{Daniel Ritchie}.}
  \bibinfo{year}{2024}\natexlab{}.
\newblock \showarticletitle{PossibleImpossibles: Exploratory Procedural Design
  of Impossible Structures}. In \bibinfo{booktitle}{\emph{Computer Graphics
  Forum}}, Vol.~\bibinfo{volume}{43}. Wiley Online Library,
  \bibinfo{pages}{e15052}.
\newblock


\bibitem[Linton et~al\mbox{.}(2023)]%
        {linton2023new}
\bibfield{author}{\bibinfo{person}{Paul Linton}, \bibinfo{person}{Michael~J
  Morgan}, \bibinfo{person}{Jenny~CA Read}, \bibinfo{person}{Dhanraj
  Vishwanath}, \bibinfo{person}{Sarah~H Creem-Regehr}, {and}
  \bibinfo{person}{Fulvio Domini}.} \bibinfo{year}{2023}\natexlab{}.
\newblock \bibinfo{title}{New approaches to 3D vision}.
\newblock , \bibinfo{numpages}{20210443}~pages.
\newblock


\bibitem[Lipman et~al\mbox{.}(2004)]%
        {lipman2004differential}
\bibfield{author}{\bibinfo{person}{Yaron Lipman}, \bibinfo{person}{Olga
  Sorkine}, \bibinfo{person}{Daniel Cohen-Or}, \bibinfo{person}{David Levin},
  \bibinfo{person}{Christian Rossi}, {and} \bibinfo{person}{Hans-Peter
  Seidel}.} \bibinfo{year}{2004}\natexlab{}.
\newblock \showarticletitle{Differential coordinates for interactive mesh
  editing}. In \bibinfo{booktitle}{\emph{Proceedings Shape Modeling
  Applications, 2004.}} Ieee, \bibinfo{pages}{181--190}.
\newblock


\bibitem[Liu et~al\mbox{.}(2019)]%
        {liu2019softras}
\bibfield{author}{\bibinfo{person}{Shichen Liu}, \bibinfo{person}{Tianye Li},
  \bibinfo{person}{Weikai Chen}, {and} \bibinfo{person}{Hao Li}.}
  \bibinfo{year}{2019}\natexlab{}.
\newblock \showarticletitle{Soft Rasterizer: A Differentiable Renderer for
  Image-based 3D Reasoning}.
\newblock \bibinfo{journal}{\emph{The IEEE International Conference on Computer
  Vision (ICCV)}} (\bibinfo{date}{Oct} \bibinfo{year}{2019}).
\newblock


\bibitem[Livingstone(2022)]%
        {livingstone2022vision}
\bibfield{author}{\bibinfo{person}{Margaret~S Livingstone}.}
  \bibinfo{year}{2022}\natexlab{}.
\newblock \bibinfo{booktitle}{\emph{Vision and art (updated and expanded
  edition)}}.
\newblock \bibinfo{publisher}{Abrams}.
\newblock


\bibitem[Loop(1987)]%
        {Loop1987SmoothSS}
\bibfield{author}{\bibinfo{person}{Charles~T. Loop}.}
  \bibinfo{year}{1987}\natexlab{}.
\newblock \showarticletitle{Smooth Subdivision Surfaces Based on Triangles}.
\newblock
\urldef\tempurl%
\url{https://api.semanticscholar.org/CorpusID:116150707}
\showURL{%
\tempurl}


\bibitem[Ma et~al\mbox{.}(2013)]%
        {ma2013change}
\bibfield{author}{\bibinfo{person}{Li-Qian Ma}, \bibinfo{person}{Kun Xu},
  \bibinfo{person}{Tien-Tsin Wong}, \bibinfo{person}{Bi-Ye Jiang}, {and}
  \bibinfo{person}{Shi-Min Hu}.} \bibinfo{year}{2013}\natexlab{}.
\newblock \showarticletitle{Change blindness images}.
\newblock \bibinfo{journal}{\emph{IEEE transactions on visualization and
  computer graphics}} \bibinfo{volume}{19}, \bibinfo{number}{11}
  (\bibinfo{year}{2013}), \bibinfo{pages}{1808--1819}.
\newblock
\urldef\tempurl%
\url{https://cg.cs.tsinghua.edu.cn/papers/TVCG-2013-changeblindness.pdf}
\showURL{%
\tempurl}


\bibitem[Marr(1982)]%
        {marr2010vision}
\bibfield{author}{\bibinfo{person}{David Marr}.}
  \bibinfo{year}{1982}\natexlab{}.
\newblock \bibinfo{booktitle}{\emph{Vision: A computational investigation into
  the human representation and processing of visual information}}.
\newblock \bibinfo{publisher}{MIT press}.
\newblock


\bibitem[Nicolet et~al\mbox{.}(2021)]%
        {Nicolet2021Large}
\bibfield{author}{\bibinfo{person}{Baptiste Nicolet}, \bibinfo{person}{Alec
  Jacobson}, {and} \bibinfo{person}{Wenzel Jakob}.}
  \bibinfo{year}{2021}\natexlab{}.
\newblock \showarticletitle{Large steps in inverse rendering of geometry}.
\newblock \bibinfo{journal}{\emph{ACM Transactions on Graphics (TOG)}}
  \bibinfo{volume}{40}, \bibinfo{number}{6} (\bibinfo{year}{2021}),
  \bibinfo{pages}{1--13}.
\newblock


\bibitem[Okano et~al\mbox{.}(2010)]%
        {yu2010embedded}
\bibfield{author}{\bibinfo{person}{Yu Okano}, \bibinfo{person}{Shogo
  Fukushima}, \bibinfo{person}{Masahiro Furukawa}, {and}
  \bibinfo{person}{Hiroyuki Kajimoto}.} \bibinfo{year}{2010}\natexlab{}.
\newblock \showarticletitle{Embedded Motion: Generating the Perception of
  Motion in Peripheral Vision}. In \bibinfo{booktitle}{\emph{ACM SIGGRAPH ASIA
  2010 Posters}} (Seoul, Republic of Korea) \emph{(\bibinfo{series}{Sa '10})}.
  \bibinfo{publisher}{Association for Computing Machinery},
  \bibinfo{address}{New York, NY, USA}, Article \bibinfo{articleno}{41},
  \bibinfo{numpages}{1}~pages.
\newblock
\showISBNx{9781450305242}
\urldef\tempurl%
\url{https://doi.org/10.1145/1900354.1900400}
\showDOI{\tempurl}


\bibitem[Oliva et~al\mbox{.}(2006)]%
        {oliva2006hybrid}
\bibfield{author}{\bibinfo{person}{Aude Oliva}, \bibinfo{person}{Antonio
  Torralba}, {and} \bibinfo{person}{Philippe~G Schyns}.}
  \bibinfo{year}{2006}\natexlab{}.
\newblock \showarticletitle{Hybrid images}.
\newblock \bibinfo{journal}{\emph{ACM Transactions on Graphics (TOG)}}
  \bibinfo{volume}{25}, \bibinfo{number}{3} (\bibinfo{year}{2006}),
  \bibinfo{pages}{527--532}.
\newblock
\urldef\tempurl%
\url{https://stanford.edu/class/ee367/reading/OlivaTorralb_Hybrid_Siggraph06.pdf}
\showURL{%
\tempurl}


\bibitem[Owada and Fujiki(2008)]%
        {owada2008dynafusion}
\bibfield{author}{\bibinfo{person}{Shigeru Owada} {and} \bibinfo{person}{Jun
  Fujiki}.} \bibinfo{year}{2008}\natexlab{}.
\newblock \showarticletitle{Dynafusion: A modeling system for interactive
  impossible objects}. In \bibinfo{booktitle}{\emph{Proceedings of the 6th
  international symposium on Non-photorealistic animation and rendering}}.
  \bibinfo{pages}{65--68}.
\newblock


\bibitem[Paszke et~al\mbox{.}(2019)]%
        {pytorch}
\bibfield{author}{\bibinfo{person}{Adam Paszke}, \bibinfo{person}{Sam Gross},
  \bibinfo{person}{Francisco Massa}, \bibinfo{person}{Adam Lerer},
  \bibinfo{person}{James Bradbury}, \bibinfo{person}{Gregory Chanan},
  \bibinfo{person}{Trevor Killeen}, \bibinfo{person}{Zeming Lin},
  \bibinfo{person}{Natalia Gimelshein}, \bibinfo{person}{Luca Antiga},
  \bibinfo{person}{Alban Desmaison}, \bibinfo{person}{Andreas Kopf},
  \bibinfo{person}{Edward Yang}, \bibinfo{person}{Zachary DeVito},
  \bibinfo{person}{Martin Raison}, \bibinfo{person}{Alykhan Tejani},
  \bibinfo{person}{Sasank Chilamkurthy}, \bibinfo{person}{Benoit Steiner},
  \bibinfo{person}{Lu Fang}, \bibinfo{person}{Junjie Bai}, {and}
  \bibinfo{person}{Soumith Chintala}.} \bibinfo{year}{2019}\natexlab{}.
\newblock \showarticletitle{PyTorch: An Imperative Style, High-Performance Deep
  Learning Library}.
\newblock In \bibinfo{booktitle}{\emph{Advances in Neural Information
  Processing Systems 32}}. \bibinfo{publisher}{Curran Associates, Inc.},
  \bibinfo{pages}{8024--8035}.
\newblock
\urldef\tempurl%
\url{http://papers.neurips.cc/paper/9015-pytorch-an-imperative-style-high-performance-deep-learning-library.pdf}
\showURL{%
\tempurl}


\bibitem[Penrose and Penrose(1958)]%
        {penrose1958impossible}
\bibfield{author}{\bibinfo{person}{Lionel~S Penrose} {and}
  \bibinfo{person}{Roger Penrose}.} \bibinfo{year}{1958}\natexlab{}.
\newblock \showarticletitle{Impossible objects: a special type of visual
  illusion.}
\newblock \bibinfo{journal}{\emph{British Journal of Psychology}}
  (\bibinfo{year}{1958}).
\newblock


\bibitem[Penrose(1993)]%
        {Penrose1993OnTC}
\bibfield{author}{\bibinfo{person}{Roger Penrose}.}
  \bibinfo{year}{1993}\natexlab{}.
\newblock \showarticletitle{On the Cohomology of Impossible Figures}.
\newblock \bibinfo{journal}{\emph{Leonardo}}  \bibinfo{volume}{25}
  (\bibinfo{year}{1993}), \bibinfo{pages}{245--247}.
\newblock
\urldef\tempurl%
\url{https://api.semanticscholar.org/CorpusID:125905129}
\showURL{%
\tempurl}


\bibitem[Ravi et~al\mbox{.}(2020)]%
        {ravi2020pytorch3d}
\bibfield{author}{\bibinfo{person}{Nikhila Ravi}, \bibinfo{person}{Jeremy
  Reizenstein}, \bibinfo{person}{David Novotny}, \bibinfo{person}{Taylor
  Gordon}, \bibinfo{person}{Wan-Yen Lo}, \bibinfo{person}{Justin Johnson},
  {and} \bibinfo{person}{Georgia Gkioxari}.} \bibinfo{year}{2020}\natexlab{}.
\newblock \showarticletitle{Accelerating 3D Deep Learning with PyTorch3D}.
\newblock \bibinfo{journal}{\emph{arXiv:2007.08501}} (\bibinfo{year}{2020}).
\newblock


\bibitem[Reutersvärd(1934)]%
        {reutersvard_triangle_1934}
\bibfield{author}{\bibinfo{person}{Oscar Reutersvärd}.}
  \bibinfo{year}{1934}\natexlab{}.
\newblock \bibinfo{title}{First Impossible Triangle}.
\newblock
\newblock
\urldef\tempurl%
\url{https://sis.modernamuseet.se/objects/28602/}
\showURL{%
\tempurl}
\newblock
\shownote{Earliest known impossible figure, later reproduced as lithographs}.


\bibitem[Roman et~al\mbox{.}(2004)]%
        {roman2004interactive}
\bibfield{author}{\bibinfo{person}{Augusto Roman}, \bibinfo{person}{Gaurav
  Garg}, {and} \bibinfo{person}{Marc Levoy}.} \bibinfo{year}{2004}\natexlab{}.
\newblock \showarticletitle{Interactive design of multi-perspective images for
  visualizing urban landscapes}. In \bibinfo{booktitle}{\emph{IEEE
  visualization 2004}}. Ieee, \bibinfo{pages}{537--544}.
\newblock


\bibitem[S{\'a}nchez-Reyes and Chac{\'o}n(2020)]%
        {sanchez2020make}
\bibfield{author}{\bibinfo{person}{Javier S{\'a}nchez-Reyes} {and}
  \bibinfo{person}{Jes{\'u}s~M Chac{\'o}n}.} \bibinfo{year}{2020}\natexlab{}.
\newblock \showarticletitle{How to make impossible objects possible: Anamorphic
  deformation of textured NURBS}.
\newblock \bibinfo{journal}{\emph{Computer Aided Geometric Design}}
  \bibinfo{volume}{78} (\bibinfo{year}{2020}), \bibinfo{pages}{101826}.
\newblock


\bibitem[Savransky et~al\mbox{.}(1999)]%
        {savransky1999modeling}
\bibfield{author}{\bibinfo{person}{Guillermo Savransky}, \bibinfo{person}{Dan
  Dimerman}, {and} \bibinfo{person}{Craig Gotsman}.}
  \bibinfo{year}{1999}\natexlab{}.
\newblock \showarticletitle{Modeling and Rendering Escher-Like Impossible
  Scenes}. In \bibinfo{booktitle}{\emph{Computer Graphics Forum}},
  Vol.~\bibinfo{volume}{18}. Wiley Online Library, \bibinfo{pages}{173--179}.
\newblock


\bibitem[Schuster(1964)]%
        {schuster1964new}
\bibfield{author}{\bibinfo{person}{Donald~H Schuster}.}
  \bibinfo{year}{1964}\natexlab{}.
\newblock \showarticletitle{A new ambiguous figure: A threestick clevis.}
\newblock \bibinfo{journal}{\emph{The American Journal of Psychology}}
  (\bibinfo{year}{1964}).
\newblock


\bibitem[Simon(1967)]%
        {simon1967information}
\bibfield{author}{\bibinfo{person}{Herbert~A. Simon}.}
  \bibinfo{year}{1967}\natexlab{}.
\newblock \showarticletitle{An Information-processing Explanation Of Some
  Perceptual Phenomena}.
\newblock \bibinfo{journal}{\emph{British Journal of Psychology}}
  \bibinfo{volume}{58}, \bibinfo{number}{1-2} (\bibinfo{year}{1967}),
  \bibinfo{pages}{1--12}.
\newblock
\urldef\tempurl%
\url{https://doi.org/10.1111/j.2044-8295.1967.tb01051.x}
\showDOI{\tempurl}
\showeprint{https://bpspsychub.onlinelibrary.wiley.com/doi/pdf/10.1111/j.2044-8295.1967.tb01051.x}


\bibitem[Sorkine(2006)]%
        {sorkine2006differential}
\bibfield{author}{\bibinfo{person}{Olga Sorkine}.}
  \bibinfo{year}{2006}\natexlab{}.
\newblock \showarticletitle{Differential representations for mesh processing}.
  In \bibinfo{booktitle}{\emph{Computer Graphics Forum}},
  Vol.~\bibinfo{volume}{25}. Wiley Online Library, \bibinfo{pages}{789--807}.
\newblock


\bibitem[Sugihara(1997)]%
        {sugihara1997three}
\bibfield{author}{\bibinfo{person}{Kokichi Sugihara}.}
  \bibinfo{year}{1997}\natexlab{}.
\newblock \showarticletitle{Three-dimensional realization of anomalous
  pictures—An application of picture interpretation theory to toy design}.
\newblock \bibinfo{journal}{\emph{Pattern Recognition}} \bibinfo{volume}{30},
  \bibinfo{number}{7} (\bibinfo{year}{1997}), \bibinfo{pages}{1061--1067}.
\newblock


\bibitem[Taylor(2020)]%
        {taylor2020modeling}
\bibfield{author}{\bibinfo{person}{Benjamin~Adam Taylor}.}
  \bibinfo{year}{2020}\natexlab{}.
\newblock \bibinfo{booktitle}{\emph{Modeling and Rendering Three-Dimensional
  Impossible Objects}}.
\newblock \bibinfo{publisher}{Bangor University (United Kingdom)}.
\newblock


\bibitem[Térouanne(1984)]%
        {TEROUANNE1984105}
\bibfield{author}{\bibinfo{person}{E. Térouanne}.}
  \bibinfo{year}{1984}\natexlab{}.
\newblock \showarticletitle{Impossible Objects and Inconsistent
  Interpretations}.
\newblock In \bibinfo{booktitle}{\emph{Trends in Mathematical Psychology}},
  \bibfield{editor}{\bibinfo{person}{E.~Degreef} {and} \bibinfo{person}{J.~{Van
  Buggenhaut}}} (Eds.). \bibinfo{series}{Advances in Psychology},
  Vol.~\bibinfo{volume}{20}. \bibinfo{publisher}{North-Holland},
  \bibinfo{pages}{105--120}.
\newblock
\showISSN{0166-4115}
\urldef\tempurl%
\url{https://doi.org/10.1016/S0166-4115(08)62082-8}
\showDOI{\tempurl}


\bibitem[Wang et~al\mbox{.}(2020)]%
        {wang2020toward}
\bibfield{author}{\bibinfo{person}{Xi Wang}, \bibinfo{person}{Zoya Bylinskii},
  \bibinfo{person}{Aaron Hertzmann}, {and} \bibinfo{person}{Robert Pepperell}.}
  \bibinfo{year}{2020}\natexlab{}.
\newblock \showarticletitle{Toward quantifying ambiguities in artistic images}.
\newblock \bibinfo{journal}{\emph{ACM Transactions on Applied Perception
  (TAP)}} \bibinfo{volume}{17}, \bibinfo{number}{4} (\bibinfo{year}{2020}),
  \bibinfo{pages}{1--10}.
\newblock


\bibitem[Weber* et~al\mbox{.}(2024)]%
        {weber2023toon3d}
\bibfield{author}{\bibinfo{person}{Ethan Weber*}, \bibinfo{person}{Riley
  Peterlinz*}, \bibinfo{person}{Rohan Mathur}, \bibinfo{person}{Frederik
  Warburg}, \bibinfo{person}{Alexei~A. Efros}, {and} \bibinfo{person}{Angjoo
  Kanazawa}.} \bibinfo{year}{2024}\natexlab{}.
\newblock \showarticletitle{Toon3D: Seeing Cartoons from a New Perspective}. In
  \bibinfo{booktitle}{\emph{arXiv}}.
\newblock


\bibitem[Wu et~al\mbox{.}(2010)]%
        {wu2010modeling}
\bibfield{author}{\bibinfo{person}{Tai-Pang Wu}, \bibinfo{person}{Chi-Wing Fu},
  \bibinfo{person}{Sai-Kit Yeung}, \bibinfo{person}{Jiaya Jia}, {and}
  \bibinfo{person}{Chi-Keung Tang}.} \bibinfo{year}{2010}\natexlab{}.
\newblock \showarticletitle{Modeling and rendering of impossible figures}.
\newblock \bibinfo{journal}{\emph{ACM Transactions on Graphics (ToG)}}
  \bibinfo{volume}{29}, \bibinfo{number}{2} (\bibinfo{year}{2010}),
  \bibinfo{pages}{1--15}.
\newblock


\end{thebibliography}

\end{document}